\documentclass[prd,aps,twocolumn,floatfix]{revtex4}
\usepackage{amssymb,graphicx}
\usepackage{amsmath}
\usepackage{amsfonts}
\usepackage{epsfig}
\usepackage[usenames]{color}
\usepackage{subfigure}
\usepackage{hyperref}

\begin{document}

\title{Turbulent flows for relativistic conformal fluids in 2+1 dimensions}

\author{
Federico Carrasco$^{1,2}$,
Luis Lehner$^{2,3}$,
Robert C. Myers$^{2}$,\\
Oscar Reula$^{1}$,
Ajay Singh$^{2,4}$}
\affiliation{
$^{1}$FaMAF-UNC, IFEG-CONICET, Ciudad Universitaria, 5000, Cordoba, Argentina.\\
$^{2}$Perimeter Institute for Theoretical Physics, Waterloo, Ontario N2L 2Y5, Canada.\\
$^{3}$Department of Physics, University of Guelph, Guelph, Ontario N1G 2W1, Canada. \\
$^{4}$Department of Physics \& Astronomy and Guelph-Waterloo Physics Institute,
University of Waterloo, Waterloo, Ontario N2L 3G1, Canada}

\begin{abstract}
We demonstrate that  relativistic conformal hydrodynamics in 2+1 dimensions
displays a turbulent behaviour which cascades energy to longer wavelengths on
both flat and spherical manifolds. Our motivation for this study is to
understand the implications for gravitational solutions through the AdS/CFT
correspondence. The observed behaviour implies gravitational perturbations of
the corresponding black brane/black hole spacetimes (for sufficiently large
scales/temperatures) will display a similar cascade towards longer wavelengths.
\end{abstract}

\maketitle

\section{Introduction}
The AdS/CFT correspondence \cite{Maldacena:1997re,magoo} provides a remarkable
framework for studying certain strongly coupled gauge theories in $d$
dimensions by mapping to weakly coupled gravitational systems in  $d+1$
dimensions. Of particular interest is the relation between gauge theory plasmas
and black hole geometries~\cite{C-3,C-4,VanRaamsdonk:2008fp}. Here the
correspondence was used to show that perturbed black brane geometries at the
classical level have a dual description in terms of fluid dynamics equations
governing long-wavelength perturbations about an equilibrium state. This
fluid/gravity correspondence builds on the highly successful and ongoing program
to calculate the near-equilibrium transport coefficients for strongly coupled
plasmas using holographic techniques, e.g., \cite{early0,early1}.

This duality has been exploited in a large number of works which exploit known
gravitational behaviour to infer properties of diverse systems described by
strongly coupled field theories in the large N limit. On the hydrodynamical
front, work in~\cite{Oz:2010wz} describes how the onset of naked singularities
can be tied to finite-time blowups in hydrodynamics. Additionally, numerical
simulations are increasingly being exploited to understand isotropization and
thermalization of systems starting far from equilibrium, e.g.,
~\cite{Chesler:2010bi,Murata:2010dx,CaronHuot:2011dr,Garfinkle:2011tc,Bantilan:2012vu,Buchel:2012gw,Chernicoff:2012gu}.

In all these works, the approach has been to understand the behaviour of
relevant systems in the gravitational side of the duality and infer from it
properties of the gauge theory dynamics. In the present work, we follow the
opposite route, namely to study particular phenomena that might arise in the
field theory to understand possibly unexpected phenomena on the gravitational
side.

Our starting point is a simple and well-known observation about the behaviour
of turbulence of Newtonian fluids in two spatial dimensions. Specifically in
this case, turbulent behaviour induces an energy cascade from short to long
wavelengths e.g.,~\cite{2012PhRvE..85c6315C}. This contrasts to the standard
cascade from long to short scales characterizing turbulence in three and higher
dimensions.\footnote{The one dimensional case has been studied
in~\cite{Liu:2010jg}.} The obvious question is therefore whether this effect
appears in the relativistic conformal fluids relevant for the AdS/CFT
correspondence, and if so, what is the dual interpretation in the gravitational
theory. A positive answer would seem to distinguish AdS gravity in four
dimensions from higher dimensions in a unique way. An affirmative answer is
also particularly intriguing as the two known instabilities in AdS spacetimes,
super-radiance and the recently found `mildly-turbulent' behaviour
in~\cite{Bizon:2011gg}, both induce energy cascades from low to high
frequencies for all $d\ge3$.

To answer this question, which has also been raised
earlier\cite{VanRaamsdonk:2008fp}, we examine the behaviour of specific fluid
flows which are dual to perturbations of Schwarzschild and Kerr black holes in
$AdS_4$. We study the problem for spherical horizons in global coordinates, as
well as planar horizons described by Poincar\'{e} coordinates. Specifically, we
study the relativistic Euler equations with the particular equation of state
corresponding to a conformal fluid on a fixed $\Re \times S^2$ manifold for the
former case while $\Re \times T^2$ for the latter. In each background, we
examine perturbations of the stationary configurations dual to the
corresponding stationary black holes. We then examine the onset of turbulence
and its cascade behaviour in this setup, and compare it with the expected
behaviour for the case of incompressible non-relativistic flows with `standard'
equations of state. We find that, as in the Newtonian case, turbulence leads to
perturbations cascading from shorter to longer wavelengths for relativistic
conformal hydrodynamics in 2+1 dimensions.

This work is organized as follows: in the remainder of this section we briefly
review some well known aspects about turbulence in two and three spatial
dimensions. Section II discusses the initial configurations considered, as well
as details of our numerical implementation. Section III presents results for
the cases considered for conformal hydrodynamics on the sphere (the results on
the torus, which are qualitatively similar, are included in an appendix). We
discuss the consequences of the obtained behaviour in section IV.
(Visualizations of the fluid flows studied here can be found in~\cite{movies}).

\subsection{\textbf{Turbulence}}

Turbulence is a ubiquitous property of fluid flows observed in nature
\cite{torrid}. Qualitatively, we might describe turbulence as a flow regime
characterized by chaotic or stochastic behaviour. Most of our theoretical
understanding of turbulence comes from the study of non-relativistic
incompressible fluids. Certainly, while a full understanding of turbulence is
not yet available, some robust results do exist. Namely, for the inviscid case
in two spatial dimensions, a global regularity theorem has been proved,
together with theorems about uniqueness and existence of solutions~\cite{Sulem}
implying no singularity of the velocity field can develop in a finite time.

For the three-dimensional case it is not yet known whether the same holds true
and resolving this issue constitutes a major open problem (see
e.g.,~\cite{easy}). It is known however, that qualitatively two and three
dimensional turbulent fluid flows exhibit profound differences arising from the
existence of a key conserved quantity which has a radical effect in the fluid's
turbulent behaviour. This quantity, dubbed \textbf{\textit{enstrophy}}, --- see
discussion around \eqref{wedge} --- has been argued to imply a very different
cascade picture in two dimensions, as compared to the three-dimensional one
\cite{ Kraichnan}. Small scales will support a \textit{`direct enstrophy
cascade'} towards smaller wavelengths, with all the enstrophy dissipation
taking place on the shortest scales. The energy flux towards small scales will
then be damped and the energy will be, instead, transferred to larger scales in
an \textit{`inverse energy cascade'}. This behaviour is observed in nature,
controlled experiments and numerical simulations.

In a phenomenological theory of two-dimensional turbulence
\cite{Kraichnan,Batchelor}, the existence of two inertial ranges was pointed
out: a direct $k^{-3}$ enstrophy cascade at small wavelengths, and an inverse
energy cascade with spectrum $k^{-5/3}$, at larger scales, were predicted.

Numerical simulations have shown the emergence of strong coherent vortex
structures that dominates the flow after some time~\cite{Mcwilliams}. These
vortices emerge as anomalous fluctuations at small scales, and have a lifetime
much longer than their characteristic eddy turnover time \cite{Borue}. As the
dynamics continues, two such vortices might collide and usually merge if they rotate
in the same direction, forming
a vortex of larger scale. In the process the energy is nearly conserved, so it
acts as the mechanism of inverse energy cascade. The statistical distribution
of vortices over scales leads to an energy spectrum  $k^{-3}$ which is much
steeper than the originally expected $k^{-5/3}$, and vortices were recognized
as the fluctuations responsible of \textit{intermittency} and possible
anomalous dimensions \cite{Benzi-2}.

The emergence and dynamics of vortices still pose a number of difficult
questions, and the relevance of the initial data and external forcing is not
yet fully understood. An exciting possibility, of course, would be that
holographic studies might shed new light on this problem from a fundamentally
different point of view.


\section{Equations, rationale and numerical implementation}
Our goal is to study the dynamics dual to a perturbed black hole in $3+1$
dimensions from the global point of view, so we consider the field theory in
$2+1$ dimensions on the sphere (or a torus, see Appendix A). To do so, and in
order to deal with a well posed problem, we restrict to the zeroth order
expansion of the theory, i.e., to the equations determined by the conservation
of the stress energy tensor of a perfect fluid. One reason for this choice is
that the inclusion of viscous terms would yield an acausal system of
equations~\cite{Muller:1967zza,Israel:1979wp,Geroch1990,Geroch1991} (for a
recent discussion in the holographic context see
e.g.,~\cite{Buchel:2009tt}).\footnote{This issue does not arise in the
Newtonian context and a holographic scenario realizing this limit has been
discussed in~\cite{Bredberg:2011jq}. Consequently, the well known turbulent
behaviour of Newtonian fluids is obtained in this case.} Hence properly
incorporating these effects poses a serious complication for the corresponding
simulations.

At first sight, working with perfect fluid equations would seem to be a severe
limitation for our study and the conclusions that can be drawn from this work.
However, recall that hydrodynamics treats the conservation of a stress-energy
tensor in a gradient expansion, e.g., \cite{Baier:2007ix,paul}. For the
conformal case, no intrinsic scales appear in defining the fluid and hence the
temperature naturally controls the equation of state and all of the
non-vanishing transport parameters. Hence to the stress tensor for a conformal
fluid, in $d$ spacetime dimensions, takes the following form
\begin{equation}
T_{\mu\nu} = \alpha\, T^d \left( (g_{\mu\nu} + d\, u_\mu u_\nu) -
\frac{d\,\beta}{T} \sigma_{\mu\nu}
+ {\cal O}(T^{-2})  \right)
\label{fluidexp2}
\end{equation}
where $\sigma_{ab}$ describes the first-order viscous contribution and
$\alpha,\ \beta$ are dimensionless coefficients which characterize the fluid.
From this expression, we see that the viscous terms can be arbitrarily
suppressed by increasing the temperature. To be more precise, if the
characteristic scales $L$ controlling the flow are kept fixed, i.e., the size
of the sphere (or torus) in the present case, the gradient expansion becomes an
expansion in $1/LT$ and the higher order terms are suppressed by setting $LT\gg
1$. Thus the perfect fluid can be seen as an arbitrarily good approximation in
this regime. Further, as we discussed above, turbulence is expected to generate
a cascade from short to long scales for fluids in two spatial dimensions. In
fact, our simulations will confirm this expectation and so the interesting
dynamics here indeed progresses mainly towards longer and longer scales. Hence
the viscous and higher order terms should not play a significant role in the
observed behaviour.

To further support the conclusion that the viscous contributions are
insignificant, we will consider two ways of incorporating effects of these
terms: (i) through the addition of an artificial viscosity (as defined in
~\cite{Kreiss,Reula}) and (ii) by comparing results for higher temperatures,
where the viscous and higher terms become less relevant (as described above),
and showing that the observed turbulent behaviour remains unchanged both
qualitatively and quantitatively, e.g., see figure \ref{T-comp}. Notice that one
could add terms up to second order in the derivative expansion (third order in
derivatives) as suggested in~\cite{Baier:2007ix} (which provides a modern
extension of the method suggested in~\cite{Israel:1979wp}). This approach
effectively amounts to controlling and reducing gradients in the solution
through a dynamical equation governing second order gradients of the flow. As
we will see, the natural evolution of the system to longer wavelengths renders
these approaches unnecessary.

In what follows we describe the system of equations considered, discuss useful
monitoring quantities --- including the conserved quantity associated to the
enstrophy ---  as well as a brief summary of stationary solutions inferred from
the dual gravitational picture.

\subsection{\textbf{Evolution Equations}}

The system follows from the local conservation of the stress-energy tensor of a perfect fluid,
\begin{equation}
 T^{\mu \nu} = (\rho + p)\, u^{\mu}\, u^{\nu} + p\, g^{\mu \nu} \, ,
\label{Tab}
 \end{equation}
along with the condition of conformal invariance that requires the stress
tensor to be traceless, i.e., $ T^{\mu}_{\mu} = 0 $. This latter condition
fixes the equation of state as
\begin{equation}
 p = \rho/2 \, ,
\end{equation}
for a conformal fluid in 2+1 dimensions --- compare with \eqref{fluidexp2} for
general dimensions.

The set of dynamical variables we have chosen to evolve the system are:  $\{
\tilde{\rho}, u_{i} \}$, where $ \tilde{\rho} \equiv log(\rho^{1/3}) $. The
remaining component of the three-velocity, $u^0 $, which can be identified with
the Lorentz factor $\gamma$ ($u^0 \equiv \gamma = (1 - v^2)^{-1/2}$), is
obtained at each time-step through the normalization condition $u_{\mu} u^{\mu}
= -1 $. In equilibrium, the temperature $T$ of the system is related with the
above variables by $\rho = T^3$, or equivalently, $T =
e^{\tilde{\rho}}$.\footnote{Here, we have dropped the constant $\alpha$
appearing in \eqref{fluidexp2}, which gives an overall normalization factor in
the energy density. Including this factor would simply produce an
inconsequential constant shift of $\tilde{\rho}$.}

The system of evolution equations obtained for our dynamical variables then
reads,
\begin{eqnarray}
\partial_t \tilde{\rho} & = & \frac{1}{\alpha \gamma} \left\lbrace z {\cal D} \tilde{\rho} +   z \vartheta - \frac{1}{2}{\cal D} z \right\rbrace \, ,  \\
\partial_t u_i & = & \frac{1}{\alpha \gamma} \left\lbrace  \left[ \alpha \partial_i  + u_i {\cal D} \right]  \tilde{\rho} + \alpha {\cal D} u_i - z u_i \vartheta  + \frac{u_i}{2} {\cal D} z \right\rbrace
\end{eqnarray}
The spatial derivatives along the flow are denoted by ${\cal D} \equiv
u^{k}D_{k} $, while $ \vartheta\equiv D_{k} u^{k} $ is the spatial
divergence and $D_k$ is the covariant derivative associated with the
spatial metric $h_{ij}$. We also have defined, $ z\equiv \gamma^{2} $ and $
\alpha \equiv 1 + \gamma^2$.

\subsection{\textbf{Equilibrium Configurations}}
Equilibrium configurations for the toroidal case are simply constant
temperature/constant velocity fluid flows and we adopt conditions as described
in the appendix. The equilibrium states for conformal fluids on the two-sphere
can be obtained straightforwardly by requiring no entropy production within the
first sub-leading (dissipative) order in the fluid expansion at equilibrium.
This imposes a restrictive condition for the shear viscosity (i.e.,
$\sigma_{ab} = 0$). A simple family of such configurations corresponds to
rigidly rotating fluids, which in our variables are given by
\begin{equation}
 u^{\theta} = 0 \qquad ,\qquad  u^{\phi} = \gamma\, \omega_{0} \qquad ,\qquad  T = \gamma\, \tau
 \label{rigid}
\end{equation}
where $ \gamma = (1 - \omega_{0}^2\, sin^{2}(\theta))^{-1/2} $ is the Lorentz
factor. The two (constant) parameters characterizing these solutions are: the
$\omega_{0}$, the angular rotation rate, and $\tau$, the local temperature
measured by co-moving observers. Note
that implicitly, we have set the radius of the $S^2$ to one here, i.e., the
proper length of the equator, $\theta=\pi/2$, is simply $2\pi$.

The thermodynamics and the local stress tensor of these solutions has been
found to be in precise agreement with thermodynamics and boundary stress tensor
of the spinning black holes \cite{C-2}. In this reference, the authors compare
conformal fluids on spheres of arbitrary dimensions with large rotating black
holes on $AdS$ spaces. First, global thermodynamical quantities are compared.
Then, appealing to the duality, a comparison is made between the local stress
tensor of the fluid configuration and the boundary stress tensor for the most
general rotating black hole in $AdS_{d+1}$, as given in ~\cite{Gibbons200549}. The
relevant black hole solution for our case, the one corresponding to Kerr
$AdS_4$, is labeled by two parameters $a$ and $r_{+}$, related with the angular
momentum (per unit mass) and the horizon radius of the black hole,
respectively. Perfect agreement was found between these two theories in the
large $r_{+}$ limit, upon the following identifications \cite{C-2}:
\begin{eqnarray}
 \omega_0 & \longleftrightarrow & a\,, \\
 \tau & \longleftrightarrow & \frac{3\,r_+}{4\pi}\,.
\end{eqnarray}
Thus, the static configuration with $\omega_{0}=0$ is the fluid dual to $AdS_4$
Schwarzschild black hole, and the rotating fluid configurations ($\omega_{0}
\neq 0$) are dual to the $AdS_4$ Kerr geometry.

\subsection{\textbf{Conservation Laws}}
Conserved quantities are invaluable tools to analyze the behaviour of any given
system and, as discussed, their existence can imply a particular behaviour of
the fluid flow. We here discuss these quantities which we monitor in our
efforts to understand the behaviour of the system, in particular the phenomenon
of turbulence. In particular, given a conserved current $J^{\mu}$ (i.e.,
$\nabla_{\mu} J^{\mu} = 0$), the existence of a conserved charge immediately
follows,
\begin{equation}
 \int_{S^2} J^{0} d\Sigma = const \, .
\end{equation}
Hence in the following, we identify various conserved currents and charges for
the present system.

Given a killing vector field $K_{\mu} $ and the conserved stress tensor, the
current $J^{\mu} \equiv T^{\mu \nu} K_{\nu} $, is automatically conserved.
Thus, from the killing vectors, $ \xi \sim \frac{\partial}{\partial t} $
 and $ \bf{\zeta} \sim \frac{\partial}{\partial \phi} $,
we construct the first two conserved quantities that are identified with the
total \textbf{energy} and \textbf{angular momentum} of the fluid, respectively:
\begin{eqnarray}
 E &\equiv&  \frac{1}{2} \int_{S^2} \rho (3 \gamma^2 - 1) d\Sigma \, , \\
 L_\phi &\equiv&  \frac{3}{2} \int_{S^2} \rho \gamma u^{\phi} d\Sigma \, .
\end{eqnarray}

From thermodynamical considerations, one knows that the local entropy $ s\equiv
\rho^{2/3}$, is conserved along the flow direction (in a perfect fluid). Thus,
the current $ J^{\mu} = su^{\mu} $ is conserved and total \textbf{entropy}
conservation follows,
\begin{equation}
 S \equiv  \int_{S^2} \rho^{2/3} \gamma d\Sigma \, .
\end{equation}

Another quantity of interest is the \textbf{vorticity} which arises as a purely
geometric property (i.e., independent of the equations of motion). It is
constructed by taking the exterior derivative to the flow velocity, resulting
in the vorticity two-form $ \omega_{\mu \nu} \equiv \partial_{[\mu} u_{\nu]} $.
In 2+1 dimensions, one can define a naturally conserved current just by taking
its Hodge dual, i.e., we define $W^{\alpha} = \epsilon^{\alpha \mu \nu}
\omega_{\mu \nu} $. The total `circulation', which remains constant throughout
the evolution, is then
\begin{equation}
 C \equiv \int_{S^2} W^0 d\Sigma =
 \int_{S^2}  {\epsilon}^{0ij}\omega_{ij}\, d\Sigma \, .
\end{equation}
However, note that the conservation law is `topological' in this case and so
this conserved charge actually vanishes since $C=-\int_{S^2}\mathbf{du}=0$.

Finally, we define the \textbf{enstrophy}, whose conservation plays a crucial
role on two-dimensional turbulence and its \textit{inverse cascade} scenario.
This quantity, for incompressible non-relativistic fluids, is just the integral
of the square vorticity field. Carter \cite{ Carter} has
generalized the concept for relativistic fluids in three spatial dimensions,
and we extend it here for two-dimensional relativistic conformal fluids.

Let us begin with $\nabla_{\mu} T^{\mu \nu} = 0$ with the stress tensor given
in \eqref{Tab} for a conformal fluid in $d$ spacetime dimensions (i.e.,
$p=\rho/(d-1)$. For convenience we project these conservation equations along
and orthogonal to the flow velocity,
\begin{eqnarray}
u^{\mu} \partial_{\mu} \rho & = & -\frac{d}{d-1}\rho (\nabla_{\mu} u^{\mu})
\equiv -\frac{d}{d-1}\rho \Theta \, , \label{parallel} \\
P^{\mu \nu} \partial_{\nu} \rho & = & -d\rho u^{\nu}\nabla_{\nu} u^{\mu}
\equiv -d\rho a^{\mu} \, ; \label{orthogonal}
\end{eqnarray}
where $\Theta$ is the (full covariant) divergence; $a^{\mu}$, the acceleration;
and $P^{\mu \nu} \equiv g^{\mu \nu} + u^{\mu} u^{\nu}$, the projector
perpendicular to $u^{\mu}$. Next, consider the two-form,
\begin{equation}
 \Omega_{\mu \nu} = \nabla_{[\mu} \rho^{1/d} u_{\nu]} \, ,
 \label{carter}
\end{equation}
which is built to satisfy the \textit{Carter-Lichnerowicz equation of motion}
\begin{equation}
 \Omega_{\mu \nu} u^{\nu} = 0 \, ,
 \label{cle}
\end{equation}
which is equivalent to \eqref{orthogonal}. Note that the inclusion of
$\rho^{1/d}$ in \eqref{carter} is crucial to produce this formulation of the
equations of motion. Then, from the \textit{Cartan Identity}, is
straightforward to show that
\begin{equation}
\textbf{\pounds}_{\lambda \mathbf{u}} \mathbf{\Omega} = \lambda \mathbf{u}\cdot
\mathbf{d\Omega} + \mathbf{d}(\lambda \mathbf{u}\cdot \mathbf{\Omega}) = 0 \, . \label{wedge}
\end{equation}
Here the first term vanishes because $\mathbf{\Omega}=\mathbf{d}
(\rho^{1/d}\mathbf{u})$ is an exact form, while the second term vanishes by the
Carter-Lichnerowicz equation \eqref{cle}. This means that the two-form
$\mathbf{\Omega}$ does not change along the flow direction.

The latter observation motivates one to look for a new conservation law,
however, we will only be able to construct the desired result by using an
identity which only holds for $d=3$, i.e., two spatial dimensions. In this
case, we write the following current: $J^{\mu} \equiv \rho^{-2/3}
(\Omega^{\alpha \beta} \Omega_{\alpha \beta}) u^{\mu} $. To establish that this
current is conserved, we make use of the identity,
\begin{equation}
 \Omega^{\mu \alpha} \Omega_{\mu \beta} = \frac{1}{2} \Omega^{2} P^{\alpha}_{\beta} \, ,
\label{key-step}
\end{equation}
where $ \Omega^2 \equiv \Omega_{\mu \nu}\Omega^{\mu \nu} $. The latter holds
for two spatial dimensions since we can write, $\Omega_{\mu \nu} =
\epsilon_{\mu \nu \alpha} l^{\alpha}$, for some arbitrary vector $l^{\alpha}$.
The condition $\Omega_{\mu \nu} u^{\nu} = 0$ then requires that $l^{\alpha}$ be
proportional to $u^{\alpha}$, i.e., $l^{\alpha} = \xi u^{\alpha}$. One can then
find the proportionality factor $\xi$ in terms of $\Omega^2 $ to show
\eqref{key-step} holds. With this relation, we have
\begin{eqnarray}
 \nabla_{\mu}J^{\mu} &=& \rho^{-\frac{2}{3}} \left\lbrace \Omega^2 \left[ \Theta -
 \frac{2 }{3\rho} u^{\mu}\partial_{\mu} \rho \right]  + u^{\mu}\partial_{\mu}\Omega^{2} \right\rbrace  \nonumber\\
                     &=& 2\rho^{-\frac{2}{3}} \left\lbrace \Theta\, \Omega^2  +
                     \Omega^{\alpha \beta}\,u^{\mu}\nabla_{\mu}\Omega_{\alpha \beta} \right\rbrace   \nonumber\\
		     &=& 2\rho^{-\frac{2}{3}} \left\lbrace \Theta\, \Omega^2  -
2 \Omega^{\mu \alpha}\Omega_{\mu \beta} (\nabla_{\alpha}u^{\beta}) \right\rbrace =0\,.   \nonumber
\end{eqnarray}
Here we have used \eqref{parallel} in the second line and \eqref{wedge} in the
third line, while the final vanishing follows from \eqref{key-step}. Therefore,
the enstrophy,
\begin{equation}
 Z \equiv \int_{S^2} \rho^{-2/3}\Omega^{2} u^{0} d\Sigma = \int_{S^2}
 ( \omega_{\mu \nu}\omega^{\mu \nu} + \frac{1}{2} a_{\mu} a^{\mu} ) \gamma d\Sigma
 \label{enstrophy}
\end{equation}
is conserved, where the first term in the expression is just the vorticity
two-form squared (as expected from the non-relativistic version), and the
second one, which involves the acceleration $a_{\mu}\equiv u^{\nu}\nabla_{\nu}
u_{\mu} $, accounts for the fact that the fluid world-lines are not necessarily
geodesics. One important point to emphasize from \eqref{enstrophy} is that the
factors of $\rho$ cancel out in the end, so that the final expression for the
enstrophy is independent of the energy density.

\subsection{Numerical approach and setup}
Our goal is to study the dynamical behaviour of a perturbed, otherwise
stationary, fluid configuration dual to the Kerr/Schwarzchild black hole in
global $AdS_4$ or the Poincar\'{e} patch. We next describe the different
components of our numerical implementation.

\subsubsection{\textbf{Initial Data}}
Initial data for perturbations of the stationary fluid configurations are
directly added to the fluid velocity such that eddys are induced. This is
straightforwardly achieved by considering perturbations which, in the co-moving
frame of the fluid, describe space varying velocities that change sign in a
smooth manner along some particular direction chosen. In the case of the torus,
this is straightforward as described in the appendix. For the spherical case,
we adopt a perturbation in the fluid velocity as:
\begin{equation}
u^{\phi} \rightarrow  u^{\phi} = \gamma\,\omega\equiv
\gamma (\omega_{0} + \delta\, \omega_{p}(\theta,\phi)) \, ,
\label{perturbation}
\end{equation}
for some small value $\delta$ and a function $\omega_{p}$ defined on the sphere
(generally chosen to be one of the spherical harmonic basis functions
$Y_{\ell}^{m}(\theta, \phi)$). In this expression, $\gamma$ is the local
Lorentz factor defined for the full angular velocity $\omega$, including the
$\omega_{p}$ contribution. For future reference, it is useful to indicate the
initial vorticity density associated to this flow,
\begin{eqnarray}
W^0(\theta) & = &
\frac{1}{\sqrt{h}} \partial_{\theta} u_{\phi} = \frac{1}{\sin(\theta)}
\partial_{\theta} \left[ \sin^{2}(\theta) \gamma (\omega_{0} + \delta
\omega_{p}(\theta,\phi)) \right] \nonumber \\
& = &  (\omega_{0} + \delta \omega_{p}) \gamma^3 \cos(\theta) (2 - \omega^2 \sin^2 (\theta))
                        \nonumber \\
     &  & + \delta \gamma^3 \sin(\theta) \partial_{\theta} \omega_{p}
                        \,. \nonumber
\end{eqnarray}
From this expression, we see that the rigid rotation component \eqref{rigid} of
the flow introduces a $Y_{1}^{0}$ component to the vorticity at order
$\omega_0$ and higher $\ell$ contributions also appear at higher orders in
$\omega_0$. Thus for sufficiently fast flows, the background contribution
renders analyzing turbulent behaviour through vorticity more delicate.

\subsubsection{\textbf{Grid Scheme}}
We now discuss details of our grid implementation and results. Since the
qualitative behaviour is the same in both topologies considered --but the
implementation is more involved in the spherical case-- from now on
we concentrate on the spherical flows to simplify the presentation and
defer details of the toroidal case to the appendix.

\begin{figure}[h!]
  \begin{center}
 \includegraphics[scale=0.4]{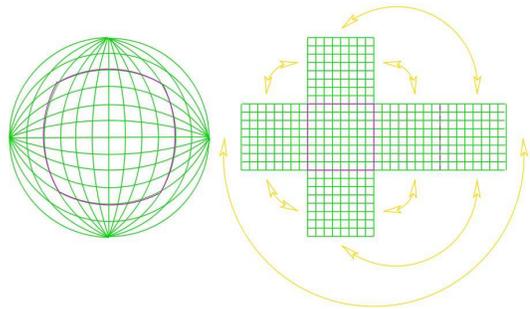}
  \caption{\textit{Cubed Sphere Coordinates.} A total of six Cartesian patches
are employed to cover the sphere. Only patch boundaries coincide at common points.}
 \label{sphere}
 \end{center}
\end{figure}

Since the topology of our computational domain is $S^2$, we employ multiple
patches to cover it in a smooth way. A convenient set of patches is defined by
the \textit{cubed sphere coordinates}. There are six patches with coordinates
projected from the sphere, and each of this patches constitute a uniform grid
--- see figure \ref{sphere}). These grids are defined in a way such that there
is no overlap and only grid points at boundaries are common to different grids
(\textit{multi-block approach}). To ensure a correct transfer of information
among the different grids we follow the technique described in~\cite{Leco},
which relies on the addition of suitable \textit{penalty terms} to the
evolution equations to preserve the energy norm through the whole sphere. This
technique `penalize' possible  mismatches between values the characteristic
fields take at interfaces and enforce consistency through suitably introduced
driving terms. Here we follow the strategy introduced in~\cite{parisireula}
which is an extension of the method introduced
in~\cite{Carpenter1999,Carpenter2001}.

Introducing coordinates $ \{ t,x,y \} $ with $(x,y)$ to label points in each Cartesian patch, the metric in each one reads,
\begin{equation}
 ds^2 = -dt^2 + \frac{1}{D^2}\left\lbrace (1 + y^2 ) dx^2 + (1 + x^2 ) dy^2 -2xy dx dy  \right\rbrace \nonumber
\end{equation}
where $ D \equiv 1 + x^2 + y^2 $. The non-vanishing Christoffel's symbols are
given by,
\begin{equation}
 \Gamma_{xx}^{x} = \frac{-2x}{D} \, \, , \, \, \Gamma_{xy}^{x} = \frac{-y}{D}  \, \, , \, \,
 \Gamma_{yy}^{y} = \frac{-2y}{D} \, \, , \, \, \Gamma_{xy}^{y} = \frac{-x}{D}   \nonumber
\end{equation}

\subsubsection{\textbf{Numerical Scheme and Stability}}

In order to construct stable finite difference schemes for our initial value problem we use the method of lines \cite{Kreiss}.
This means that we first discretize the spatial derivatives (constructing some suitable finite difference operators)  and obtain
a  system of ordinary differential equations for the grid functions.
To ensure the stability of the numerical scheme we use the \textit{energy method} described on \cite{Reula}.
We employ (fourth/second-order accurate at interior/boundary points) finite difference operators
satisfying  \textit{summation by parts} (the discrete analogue of integration by parts)
and deal with interface boundaries with appropriate penalty terms at the interfaces as described in ~\cite{ Leco}. For
the time integration, we adopt a $4^{th}$ order Runge-Kutta algorithm.

All simulations were performed using 81x81 grids for each of the six patches,
giving a total number of around 40.000 grid points to cover the entire sphere.
We have confirmed convergence through several tests, in particular we adopted
different initial data, and studied the numerical solutions with increased
resolution with each grid having  321x321, 161x161, 81x81 points (labeled by
4,2,1 respectively). With each obtained solution, denoted by $U^{(i)}(t,
\theta, \phi)$ (for $i=4,2,1 $), we calculated the \textit{convergence rate}
$p$ as, $Q(t)\equiv \frac{||U^{(4)} - U^{(2)}||_{L_{2}}}{||U^{(2)} -
U^{(1)}||_{L_{2}}} \approx 2^p$. For the cases considered, the obtained rate
was in very good agreement with the expected for resulting $3^{\text{rd}}$ order
implementation constructed.

\section{Results}

We analyze the dynamics of small perturbations around the stationary fluid
configurations duals to Schwarzschild and Kerr geometries on $AdS_4$. We
separately study these two cases, starting with a qualitative description of
the system evolutions, for generic perturbations.
We show in figures \ref{vor_sch_Y10_100} and \ref{snapshot-kerr} the sequence
on the evolution of the vorticity field for the two cases. Then, in order to
gain some insight into the turbulent behaviour and to capture the possible
cascading phenomena, we perform a spherical harmonic decomposition (up to
$\ell$=12) of the relevant fields and compute their associated power spectrum
as a function of time.

This signal processing analysis is derived simply from a generalization of
\textit{Parseval's theorem}, which states that the total power of a function
$f$ defined on the unit sphere is related to its spectral coefficients by,
\begin{equation}
 \frac{1}{4\pi} \int_{S^2} |f(\theta, \phi)|^2 d\Sigma = \sum_{\ell=0}^{\infty} C_{f} (\ell) \, ,
\nonumber
\end{equation}
\begin{equation}
 C_{f}(\ell) = \sum_{m=-\ell}^{\ell} |A_{\ell m}|^2 \, ; \nonumber
\end{equation}
where $A_{\ell m}$ are the coefficients of the expansion of $f$ in the
spherical harmonics. In figures \ref{vor_sch_Y10_100} and
\ref{vor_kerr_Y10_100}, we plot the coefficients $ C(\ell) $ as a function of
time, to analyze how the different modes behave during the evolution.

\subsection{Non-rotating case: perturbations to Schwarzschild}

\begin{figure}
\centering{
\begin{minipage}{4.2cm}
 \subfigure[$\text{  } t=0$]{\includegraphics[scale=0.2]{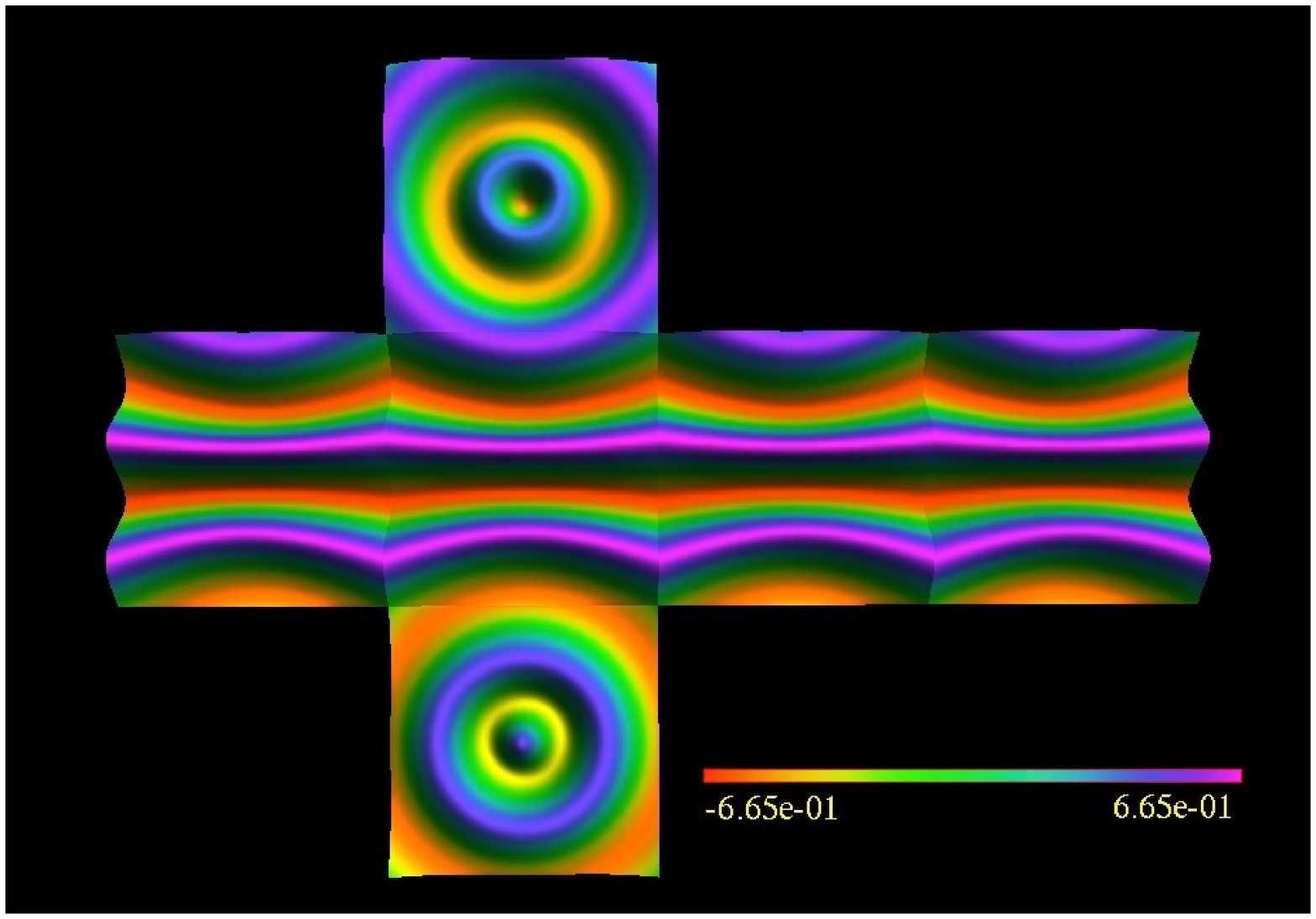}
 }
\end{minipage}
\begin{minipage}{4.2cm}
\subfigure[$\text{  } t=108$]{\includegraphics[scale=0.2]{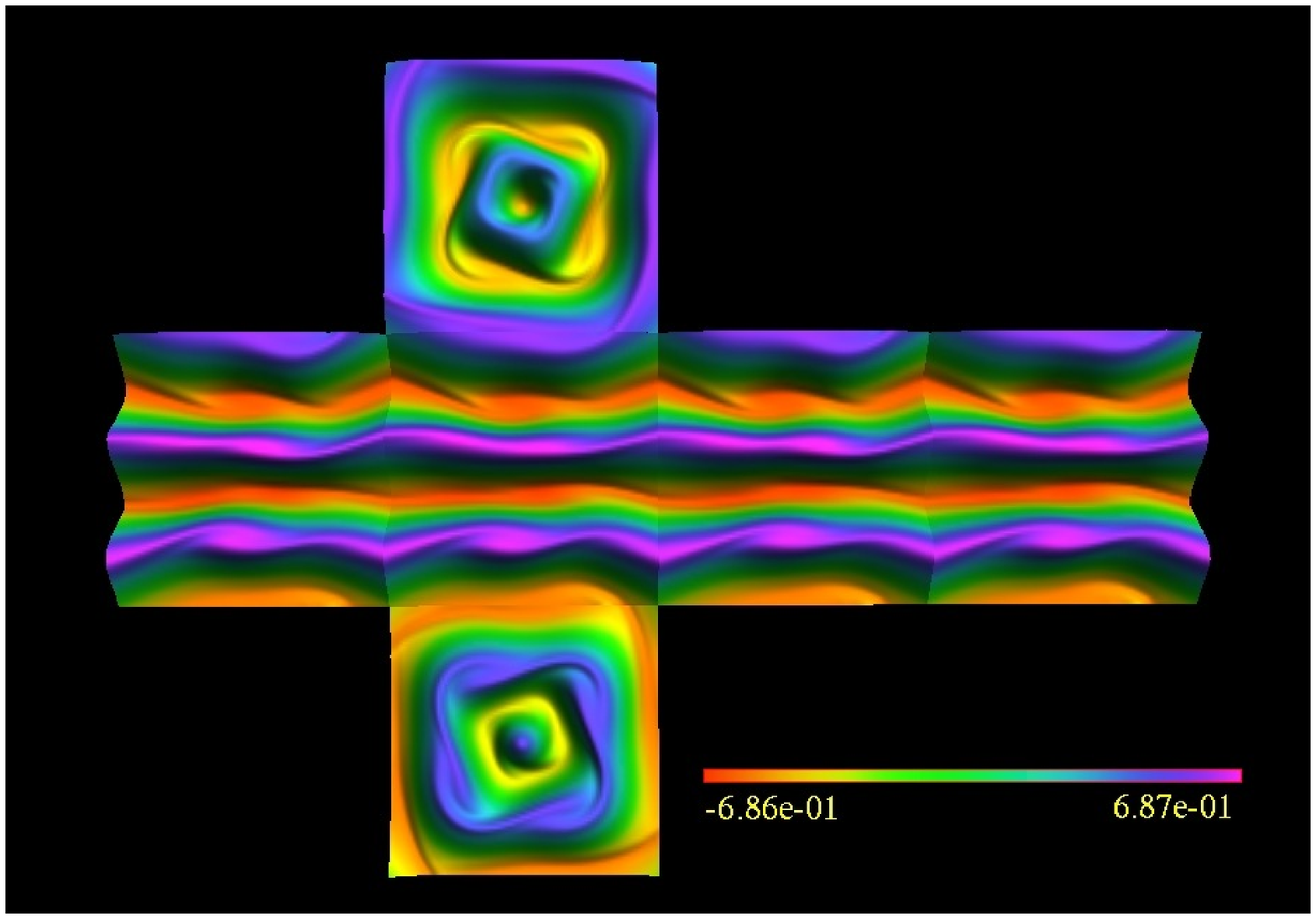}
\label{plot-1b}}
\end{minipage}}
\    \ \vfill
\centering{
\begin{minipage}{4.2cm}
\subfigure[$\text{  } t=291$]{\includegraphics[scale=0.2]{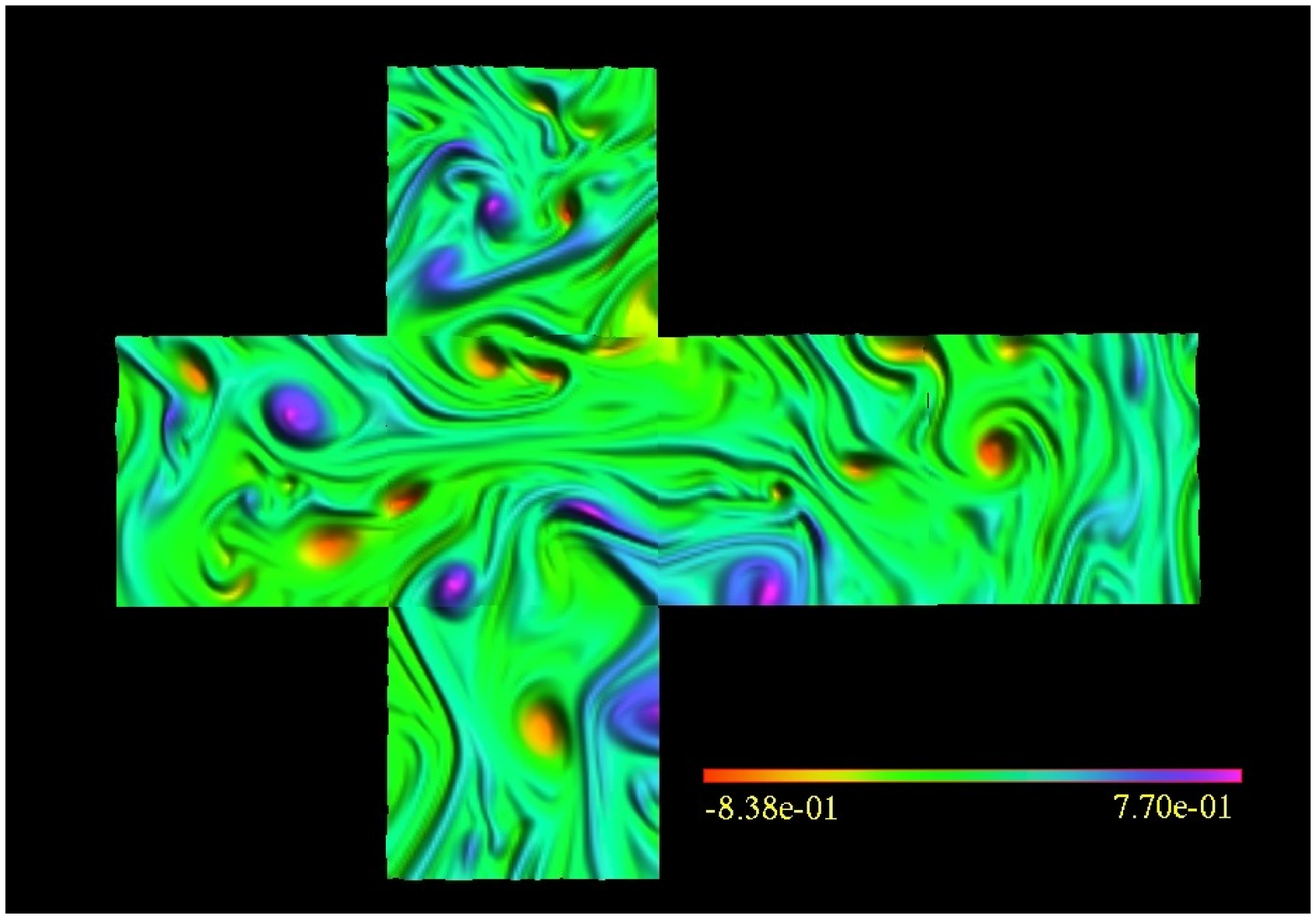}
\label{plot-1c}}
\end{minipage}
\begin{minipage}{4.2cm}
\subfigure[$\text{  } t=950$]{\includegraphics[scale=0.2]{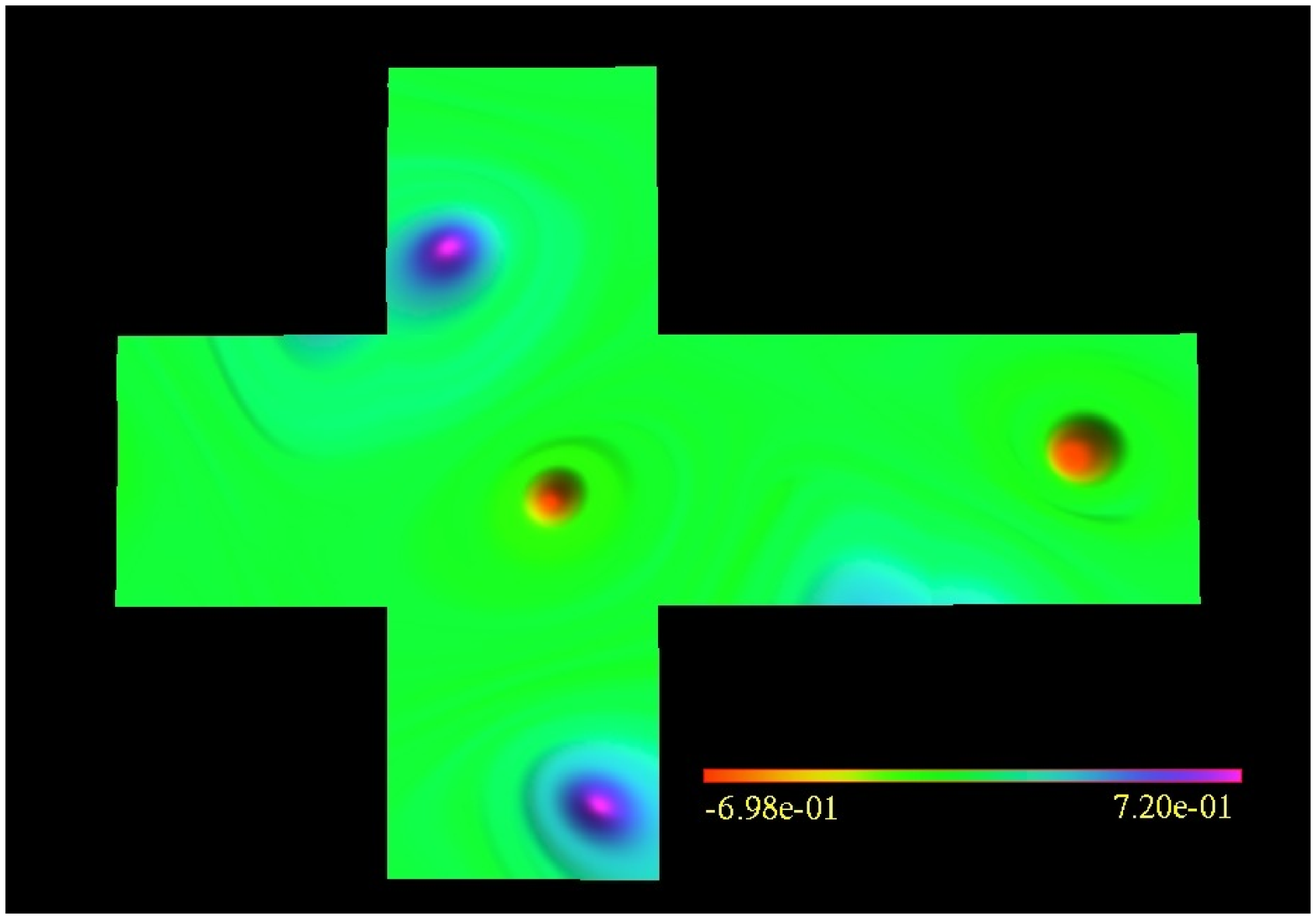}
\label{plot-1d}}
\end{minipage}

\caption{ Evolution of the vorticity field for a perturbation $\omega_{p}(\theta, \phi) = Y_{10}^{0} (\theta, \phi)$ and $\delta = 0.2$,
on Schwarzschild ($\omega_{0}$).
(a) Initial config., (b) beginning of turbulence, (c) \textit{fully developed} turbulent stage, (d) final state. }
 \label{plot-1}}
\end{figure}

\begin{figure}
\    \ \vfill
\    \ \vfill
  \begin{center}
\includegraphics[scale=0.35]{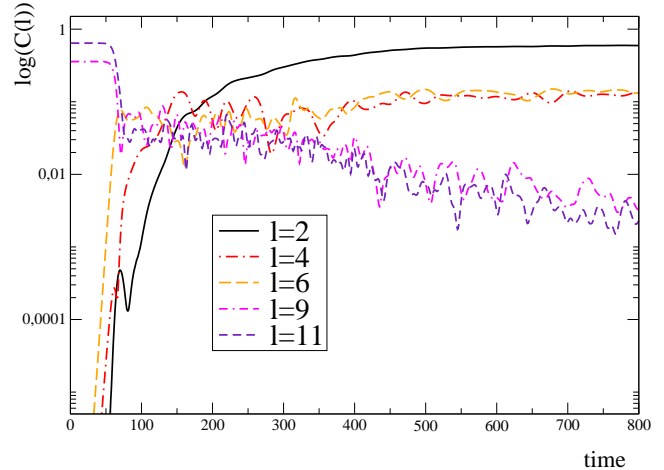}
  \caption{Relevant modes in the power spectrum of the vorticity field for a $\omega_{p}(\theta, \phi) = Y_{10}^{0} (\theta, \phi)$ and $\delta = 0.2$ perturbation,
  in the non-rotating case (at temperature $T\sim 100$).}
 \label{vor_sch_Y10_100}
 \end{center}
\end{figure}

One can basically classify the dynamical behaviour of the system into four
different stages. In figure \ref{plot-1}, we plot the vorticity field of a
representative example for each one of them. A first stage corresponds to an
initial transient period when the initial configuration seemingly remains
unchanged for some time interval. (This interval depends on the perturbation
considered, being shorter for larger $\ell$'s perturbations at a fixed value of
$\delta$, but it does not depend on the system's temperature). Closer
inspection however reveals that non-trivial dynamics begins to manifest and an
exponential growth of some modes sets in (see fig. \ref{vor_sch_Y10_100}).
Notice that the initial growth rate of these modes is approximately the same
though the higher $\ell$'s grow to larger values earlier. Since truncation
errors feed higher frequencies this behaviour is not surprising. As these modes
become large enough (roughly commensurate to the magnitude of the initial
perturbation), the original symmetry of the system gets broken and a number of
\textit{eddies} arise and move around. Such an instant might be regarded as the
beginning of turbulence (fig. \ref{plot-1b}). As the dynamics continues, the
eddies gradually turn into individual vortices and exhibit a seemingly chaotic
motion, during a stage that we will refer as \textit{fully developed
turbulence} (fig. \ref{plot-1c}). With the vortices propagating around the
sphere, encounters of same-sign vortices lead to increasingly larger vortex
structures. During this stage, governed by non-linear effects, energy transfers
from the higher $l$ modes to lower ones. The process continues until four
vortices, two of each sign, are formed (fig. \ref{plot-1d}).

The \textit{end-state} is found to be qualitatively the same in all the cases
we have considered, regardless of the initial perturbation introduced and of
the system's temperature. We have explored generic perturbations in the
velocity initial data, as described on \eqref{perturbation}: from different
single-mode perturbations (i.e., $ \omega_{p} \sim Y_{\ell}^{m}(\theta, \phi)$
for particular $\ell$ and $m$), to a random combination of all of them up to
$\ell=12$. We have further considered the inclusion of a random forcing term in
the evolution equations, by adding a term $f^{\nu}$ to the right hand side of
the stress conservation equation as: $ \nabla_{\mu} T^{\mu \nu} = f^{\nu} $.
This force is given by a random field, both in time and space (and thus, acting
on very short scales). Even in this forced case, the dynamical character and
properties of the solution remain qualitatively unchanged. This (in addition to
convergence studies) indicates that our grid structure is playing no
significant role on the obtained behaviour and on the final configuration
attained.

\begin{figure}
\centering{
\begin{minipage}{4.2cm}
 \subfigure[Vorticity]{\includegraphics[scale=0.2]{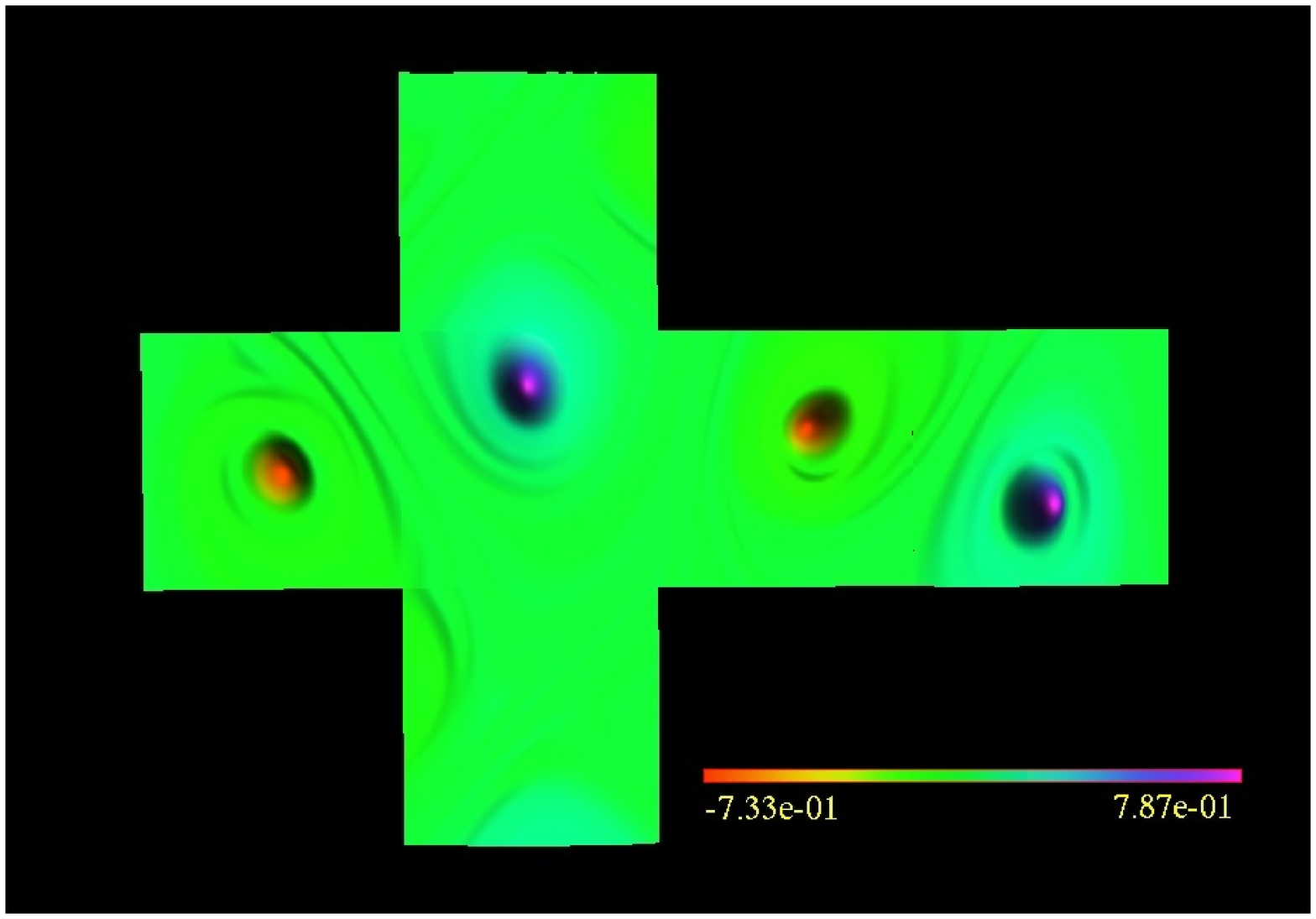}
 }
\end{minipage}
\begin{minipage}{4.2cm}
\subfigure[Temperature]{\includegraphics[scale=0.2]{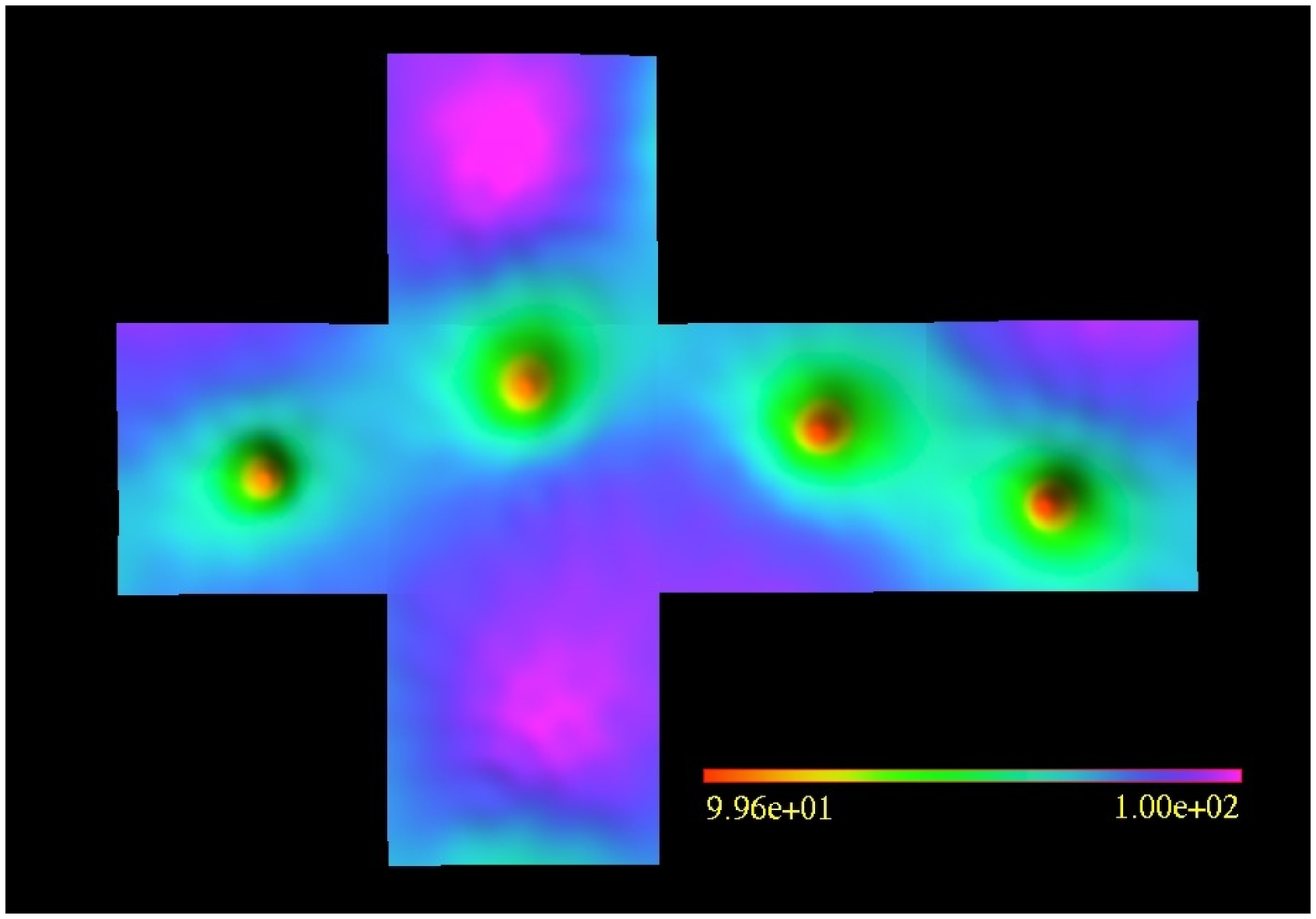}
 }
\end{minipage}}
\    \ \vfill
\centering{
\begin{minipage}{4.2cm}
\subfigure[Energy]{\includegraphics[scale=0.2]{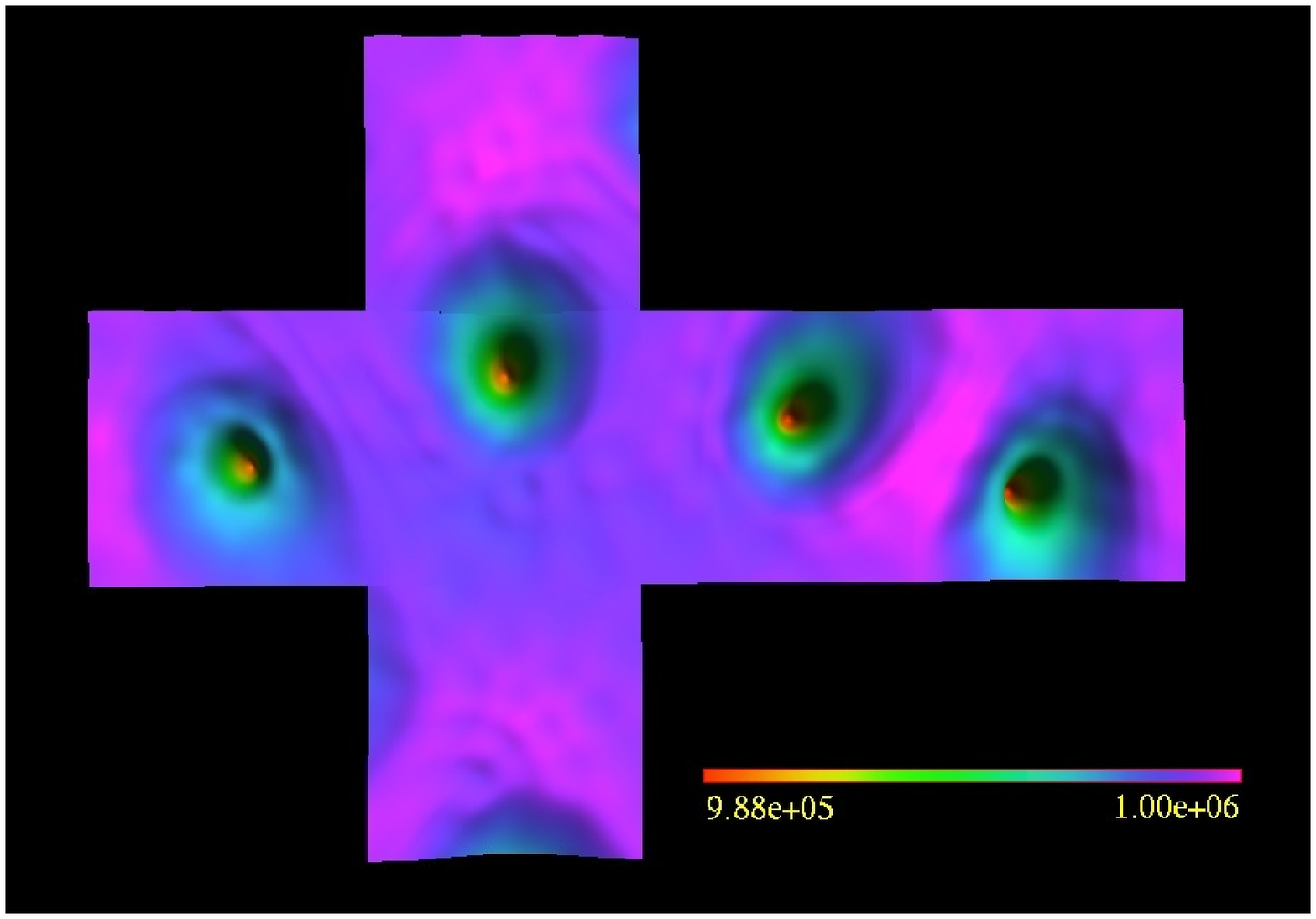}
 }
\end{minipage}
\begin{minipage}{4.2cm}
\subfigure[Enstrophy]{\includegraphics[scale=0.2]{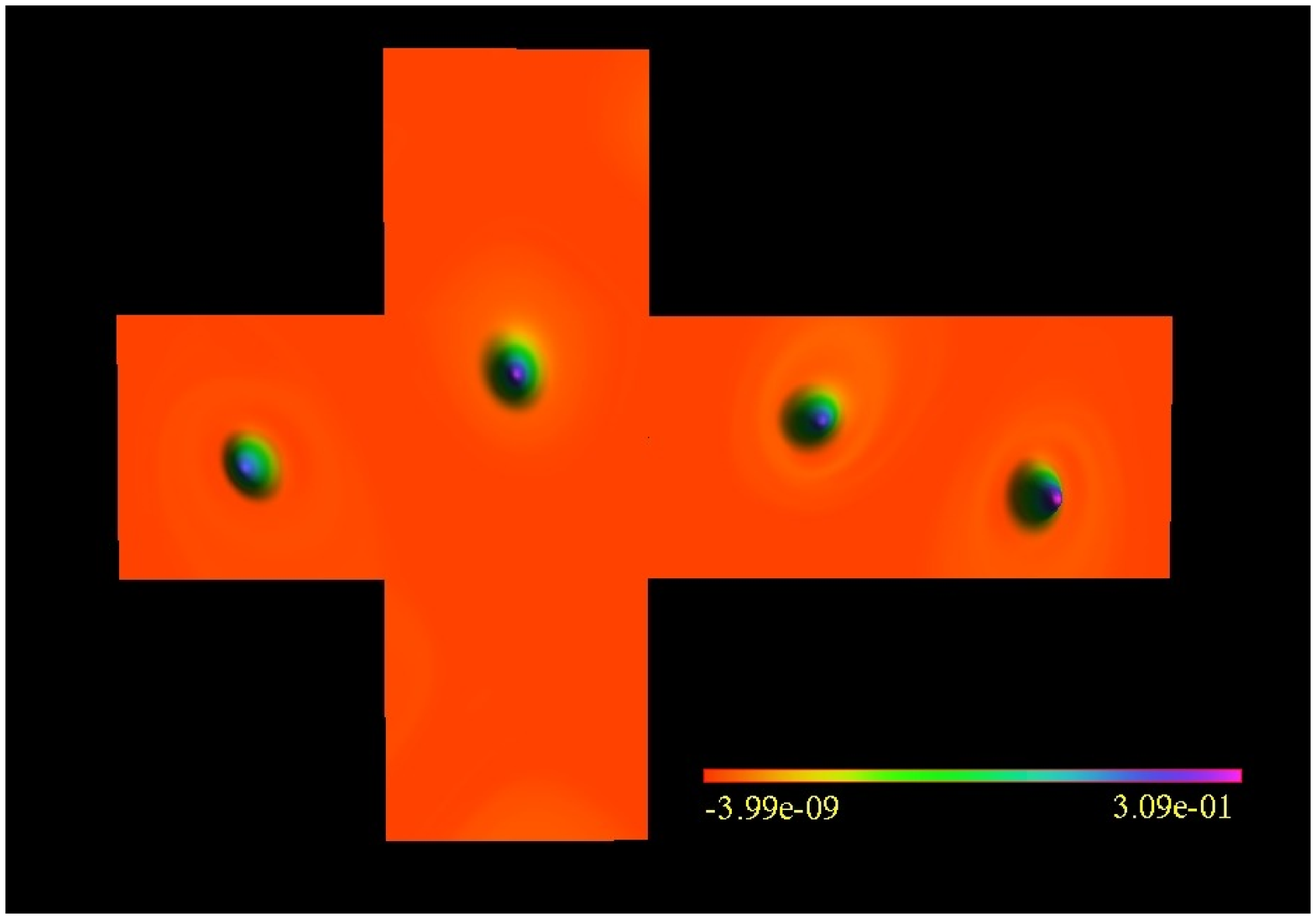}
 }
\end{minipage}

 \caption{ Late times configuration for distinct relevant fields in the
non-rotating case ($\omega_{0}=0$ and $T\sim 100$, at $t=1200$). Notice that
vortices correspond to minima in the energy. This behaviour is expected as
stable vortex structure require a larger surrounding pressure that prevents its
dispersion.}
 \label{vortices}}
\end{figure}

Main features of the long-term behaviour of the solution are illustrated in
figure \ref{vortices}. In particular, the mentioned four dominant vortices of
the resulting configuration. It can be noted that the temperature and energy
densities attain local minimums at the vortices' locations and that the
enstrophy is almost exclusively contained within them.

In figure \ref{vor_sch_Y10_100} we have displayed a few representative modes of
the vorticity power spectrum for an initial perturbation $\omega_{p}(\theta,
\phi) =  Y_{10}^{0}(\theta, \phi)$, $\delta = 0.2$ and $\tau=100$. The
coefficients $ C(\ell) $ of the spectrum are normalized with respect to the
total power and plotted (in logarithmic scale) as a function of time. In the
figure, the modes $\ell=9$ and $\ell=11$ dominate the spectrum at $t=0$ --- as
expected since the vorticity involves a derivative of the three-velocity and
the initial data for the latter has a single $\ell=10$ mode. The rest of the
modes exhibit an exponential growth on the early stages of the evolution, as
previously described. As the turbulent cascade begin, and after a short stage
in which the mode structure is rather complex, the original high-$\ell$ modes
decrease while the lower ones increase and gradually dominate the flow.
Particularly the $\ell=2$ mode in the spectrum, which represents the quadrupole
contribution, is the dominant mode at late times.

\begin{figure}
\    \ \vfill
\    \ \vfill
\centering{
\includegraphics[scale=0.35]{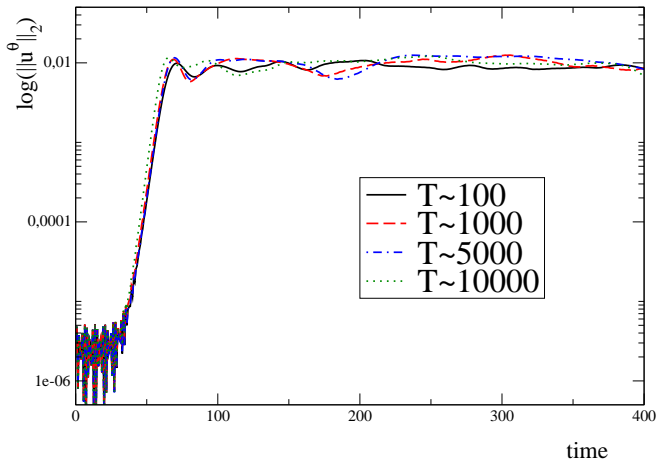}
 }

 \caption{ Dependence with temperature: logarithm of the $L_2$ norm of $u^{\theta}$
  for different temperatures, in the non-rotating case.}
 \label{T-comp}
\end{figure}

Note that the vortices in figure \ref{vortices}, i.e., the $\ell=2$ mode above,
are essentially quasi-stationary as we have neglected here the viscous (and
higher) contributions which would certainly affect the very long-time behaviour
of the solution. As we noted before, one might wonder whether the observed
turbulent phenomena, and the conclusions we can draw from this study, might be
significantly affected when such dissipative terms are taken into account.
However, recall that in the discussion around (\ref{fluidexp2}), we argued the
perfect fluid equations will give a good description when  $L T\gg 1$. We
verified the accuracy of this statement by comparing the system's dynamics at a
variety of temperatures  --- recall that the radius of the sphere is fixed to
be one --- while keeping the initial data for $u^k$ fixed. Figure \ref{T-comp}
illustrates the behaviour of the $L_2$ norm of $u^{\theta}$, i.e.,
$\left[\int_{S^2} |u^\theta|^2 d\Sigma\right]^{1/2}$. The latter is a useful
proxy for the solution's turbulent behaviour since in the absence of
turbulence, it would remain zero by symmetry considerations. As is evident in
this figure, the system's behaviour does not change as the temperature
increases. In particular, the growth rate and both onset and full development
of turbulence are essentially the same in all cases considered.




\subsection{Rotating case: perturbations to Kerr}

\begin{figure}
\centering{
\begin{minipage}{4.2cm}
 \subfigure[$\text{  } t=0$]{\includegraphics[scale=0.2]{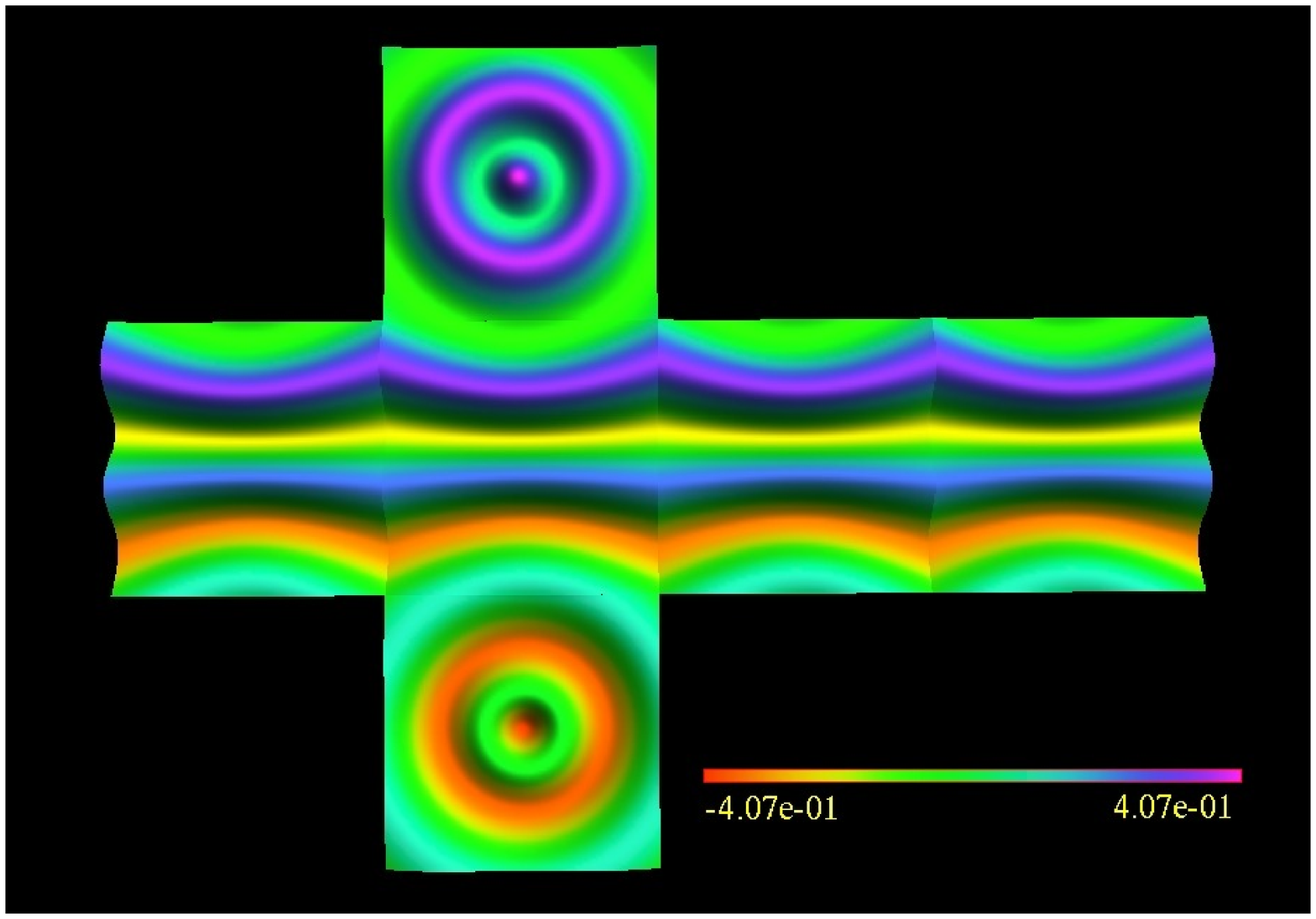}
 }
\end{minipage}
\begin{minipage}{4.2cm}
\subfigure[$\text{  } t=294$]{\includegraphics[scale=0.2]{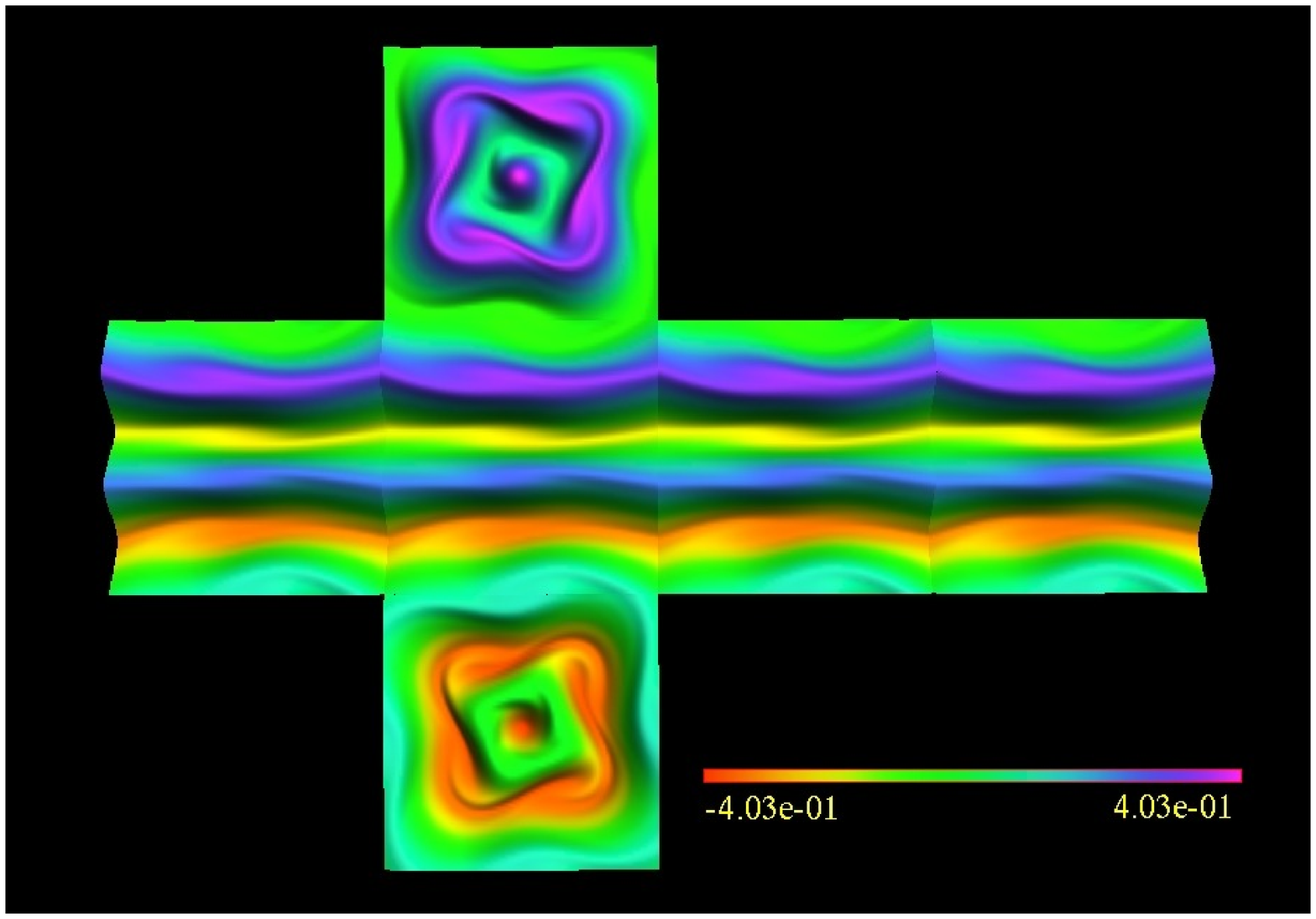}
\label{kerr-b}}
\end{minipage}}
\    \ \vfill
\centering{
\begin{minipage}{4.2cm}
\subfigure[$\text{  } t=1010$]{\includegraphics[scale=0.2]{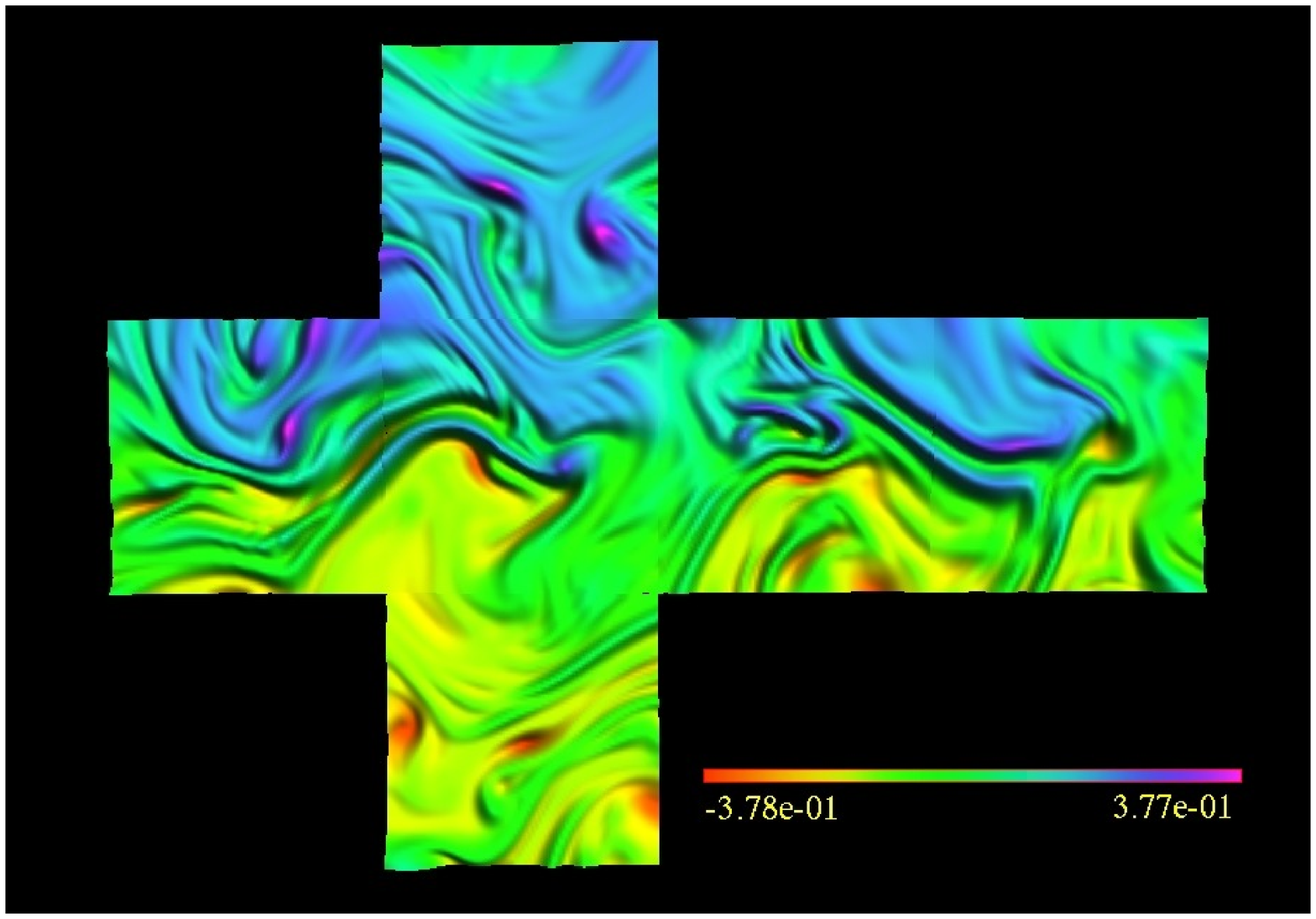}
\label{kerr-c}}
\end{minipage}
\begin{minipage}{4.2cm}
\subfigure[$\text{  } t=3500$]{\includegraphics[scale=0.2]{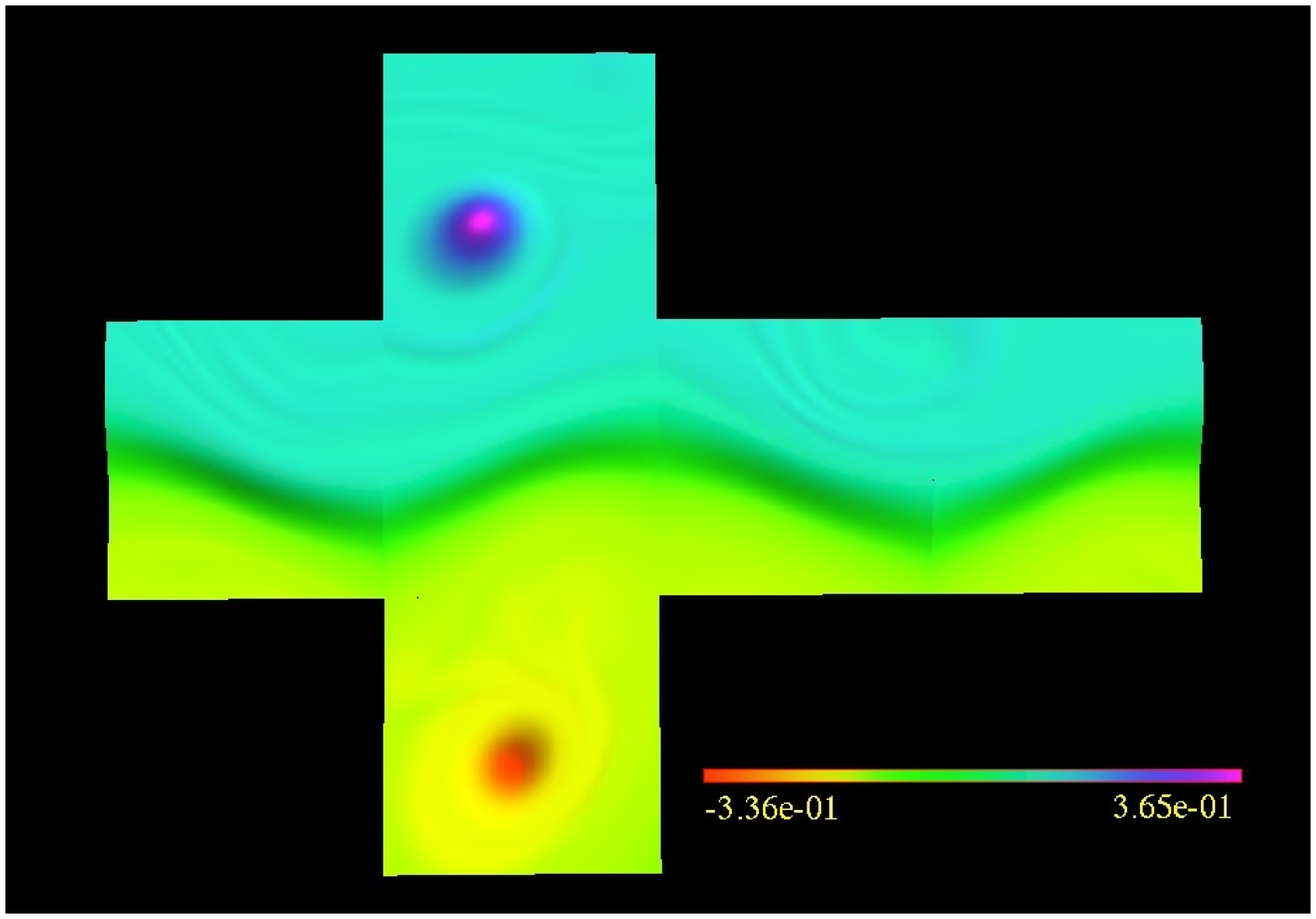}
\label{kerr-d}}
\end{minipage}

\caption{ Evolution of the vorticity field for a perturbation
$\omega_{p}(\theta, \phi) = Y_{10}^{0} (\theta, \phi)$ and $\delta = 0.08$, on
a rigid rotation with $\omega_{0} = 0.1$. (a) Initial config., (b) beginning of
turbulence, (c) \textit{fully developed} turbulent stage, (d) final state. }
 \label{snapshot-kerr}}
\end{figure}

The qualitative behaviour of the system on the rotating scenario is illustrated
in figure \ref{snapshot-kerr}, and it happens to be very similar to the
presented above for the Schwarzschild case. However, we emphasize that these
rotating solutions are not simply static solutions, i.e., $\omega_0=0$, in a
rotating frame. Recall that the rigid rotation introduces a background
vorticity field, which one finds completely dominates the long-term behaviour.
A key difference is evident immediately after the vortices are formed: the
background rotation drags them into the main rotating stream and gradually
separates them according to the direction of their rotation. If the background
rotation flows left to right as in fig. \ref{kerr-c}, clockwise rotating
vortices accumulate in the southern (lower) hemisphere while the vortices with
counter-clockwise rotation migrate towards the opposite pole. Then, the system
undergoes a merging process (of co-rotating vortices) as in the
non-rotating case, but now just one vortex of each sign remains at late times,
as in fig. \ref{kerr-d}. The initial northwards/southwards propagation of
counterclockwise/clockwise vortices can be also explained in terms of the
equal-sign mergers of vortices by regarding each of the smaller vortices generated
in the flow as interacting with two much larger vortices induced at the north
and south pole by the rigid rotation component of the flow. Note that the
\textit{`final'} two vortices are not necessarily in precise alignment with the
rotation axis, being often the case that they remain orbiting around the poles.

\begin{figure}
\centering{
\begin{minipage}{4.2cm}
 \subfigure[Vorticity]{\includegraphics[scale=0.2]{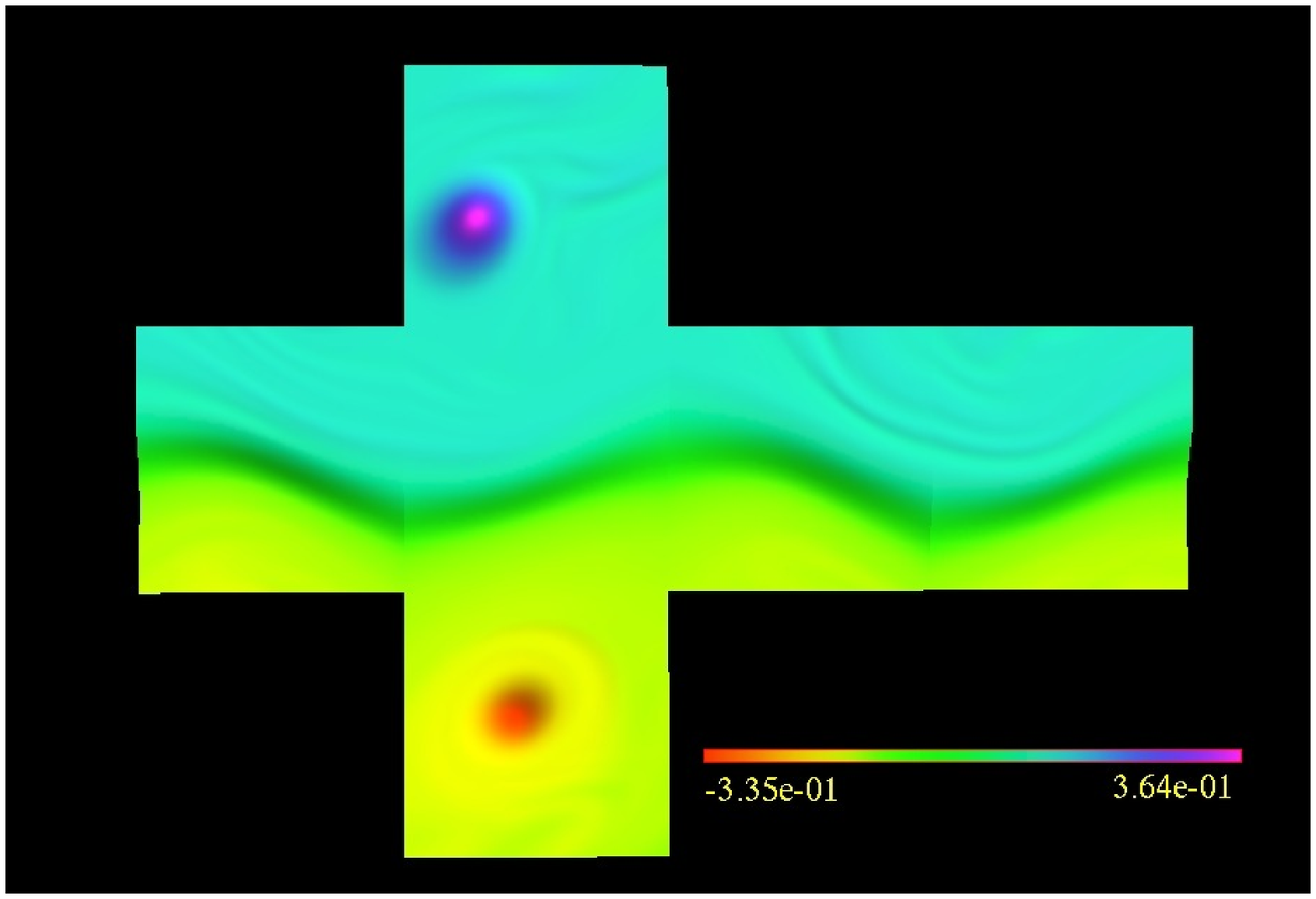}
 }
\end{minipage}
\begin{minipage}{4.2cm}
\subfigure[Temperature]{\includegraphics[scale=0.2]{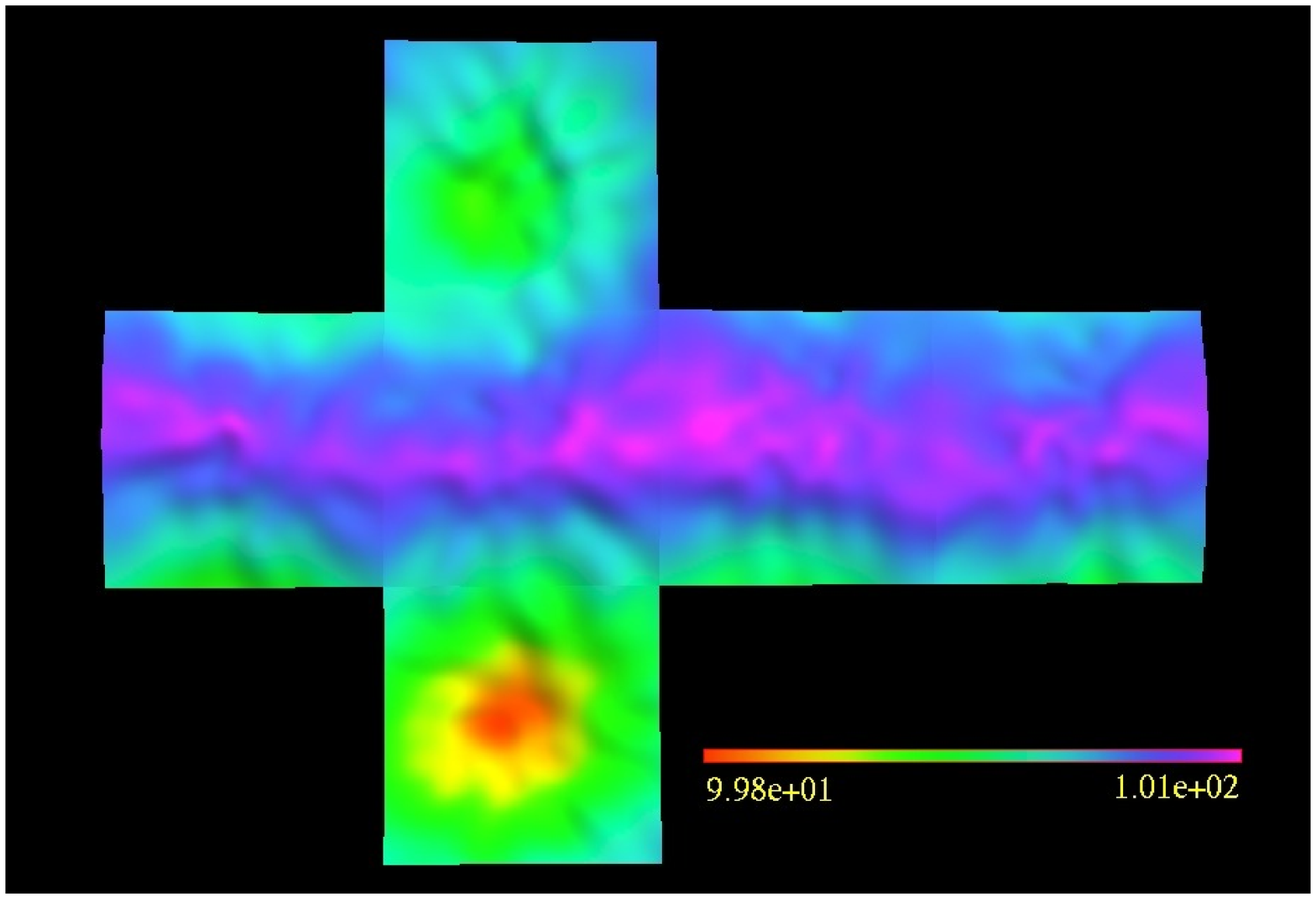}
 }
\end{minipage}}
\    \ \vfill
\centering{
\begin{minipage}{4.2cm}
\subfigure[Energy]{\includegraphics[scale=0.2]{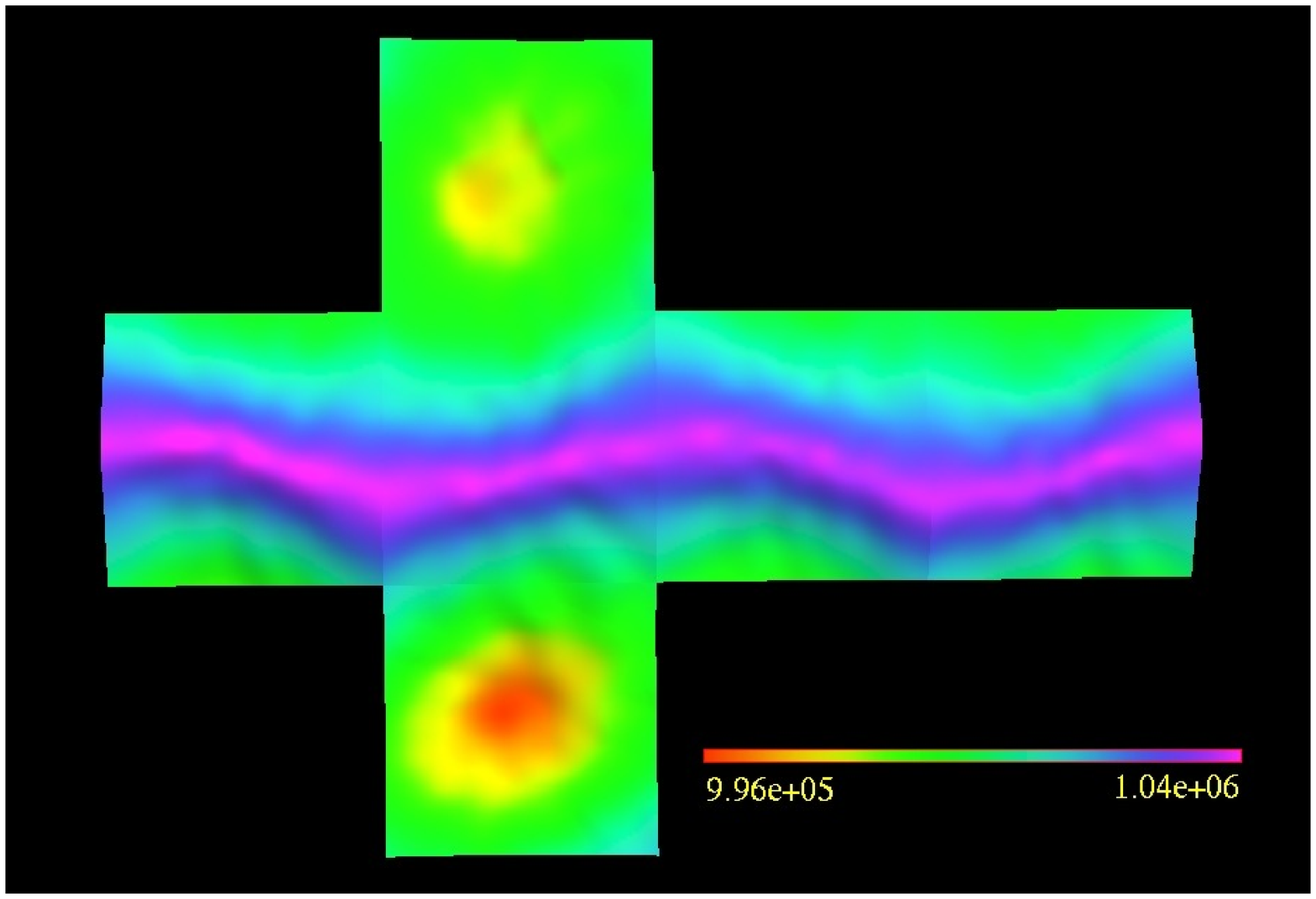}
 }
\end{minipage}
\begin{minipage}{4.2cm}
\subfigure[Enstrophy]{\includegraphics[scale=0.2]{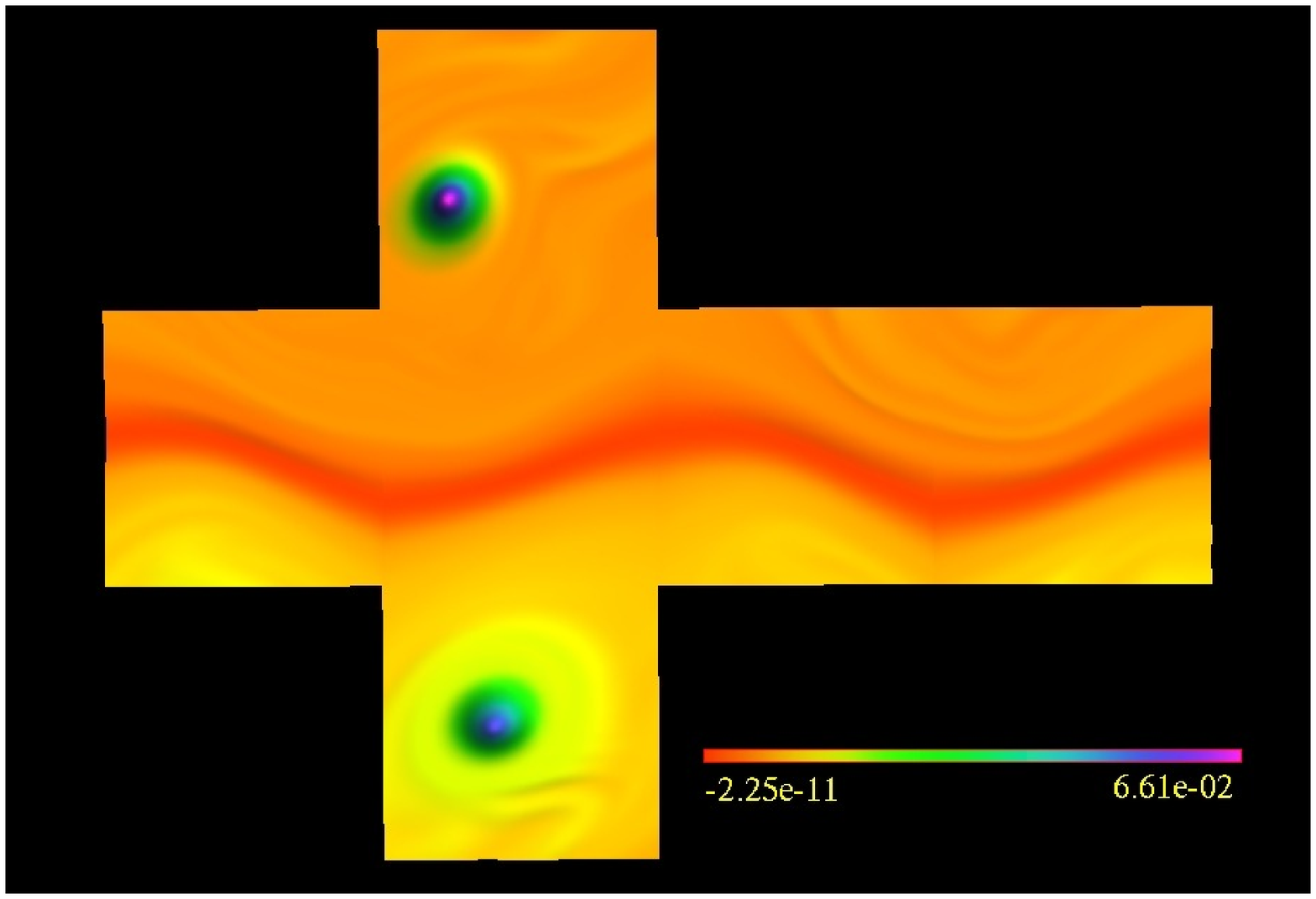}
 }
\end{minipage}

\caption{The late time configuration for distinct relevant fields in the
rotating case ($\omega_{0}=0.1$ and $T\sim 100$, at $t=3600$).  }
 \label{vortices_kerr}}
\end{figure}

Main features of the long-term behaviour of the solutions in the rotating
scenario are illustrated in figure \ref{vortices_kerr}. In particular the
mentioned two dominant vortices as they oscillate around the poles, with the
temperature and energy attaining local minima at the vortices' locations and
being concentrated near the equator where the fluid velocity is larger. The
enstrophy is again almost exclusively contained within the vortices.

\begin{figure}
\centering{
\begin{minipage}{8.cm}
 \subfigure[$\text{  } \omega_{0} = 0.1$]{\includegraphics[scale=0.3]{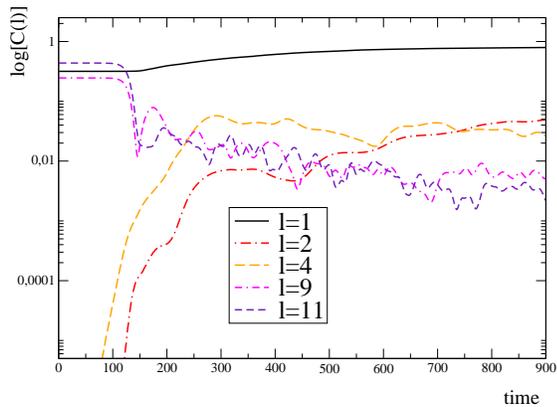}
 }
\end{minipage}
\    \ \vfill
\    \ \vfill
\    \ \vfill
\    \ \vfill
\begin{minipage}{8.cm}
\subfigure[$\text{  } \omega_{0} = 0.5$]{\includegraphics[scale=0.3]{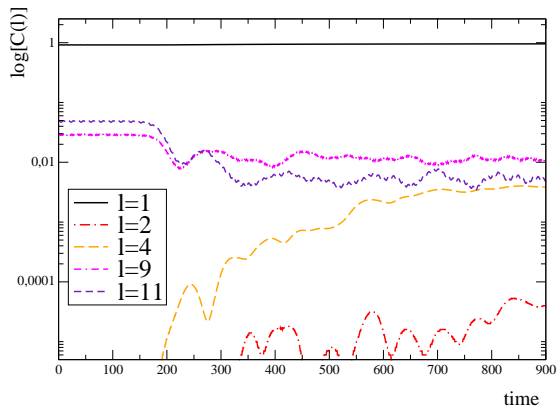}
 }
\end{minipage}}
 \caption{Relevant modes in the power spectrum of the vorticity field in the rotating case (at temperature $T\sim 100$),
  for two rotation parameters $\omega_{0}$.
  The initial perturbation is given by  $\omega_{p}(\theta, \phi) = Y_{10}^{0} (\theta, \phi)$ and $\delta = 0.08$.}
 \label{vor_kerr_Y10_100}
\end{figure}

In figure \ref{vor_kerr_Y10_100} we display representative modes of the
vorticity spectra for two different parameters of the rotating solution with
$\tau=100$ and a perturbation set by $\omega_{p}(\theta, \phi) = Y_{10}^{0}
(\theta, \phi)$ and  $\delta = 0.08$. Here, the $\ell=1$ mode represents the
background rotation contribution while the $\ell=9$ and $\ell=11$ modes are the
ones associated with the perturbation, as discussed previously. We should
recall here that the coefficients in the spectrum are normalized to unity, and
thus it becomes clear from the plots that the long term dynamics is being
completely dominated by the rotation.

For the case $\omega_{0} = 0.1$, the configuration starts with a perturbation
comparable in magnitude with the background motion. As the vortices form and
turbulence begins, the high-$\ell$'s modes decrease displaying a cascading
phenomena into lower $\ell$ modes. In particular this cascade progresses
towards the $\ell=1$ mode that eventually governs the state of the system. On
top of this rotation dominated flow, the two vortical structures mentioned
above persist and are represented in the spectrum by a combination of higher
modes, predominantly the $\ell=2$ and $\ell=4$ modes.

Interestingly, for faster initial background rotations (i.e., $\omega_{0}
\gtrsim 0.5$) while keeping the perturbation amplitude $\delta$ fixed, the flow
is already dominated by the $\ell=1$ mode from the beginning and the turbulent
stage takes longer to appear and fully develop. It seems that a strong rotating
stream in the background hinders to a certain degree the formation and merging
of vortices.

\section{Discussion and final words}
In this work we have analyzed relativistic conformal fluid flows on both $S^2$
and $T^2$ backgrounds and found first that turbulence naturally arise in these
flows and second that it gives rise to a `inverse' cascade from shorter to longer
wavelengths. These results not only extend common observations in Newtonian
hydrodynamics but also have tantalizing implications for the behaviour of
gravity in four-dimensional AdS spacetimes. Indeed, the AdS/CFT correspondence
implies that for sufficiently high temperatures/length scales, the conformal
fluid flows studied here have a dual description in terms of gravitational
perturbations on Schwarzschild or Kerr black holes in AdS$_4$, either with a
spherical or planar horizon. The turbulent behaviour observed here implies,
through the duality, that gravitational perturbations in this limit should
cascade to smaller frequencies. This constitutes a {\em prediction} obtained
within holography which had not been previously anticipated on firm 
grounds\footnote{Note however the possibility had been raised earlier 
in~\cite{VanRaamsdonk:2008fp,Evslin:2012zn}}. Indeed, previously
identified instabilities in AdS, {\em super-radiance}
(e.g.,~\cite{Winstanley:2001nx}) and {\em
`weak-turbulence'}~\cite{Bizon:2011gg,Dias:2011ss} imply a frequency shift
towards higher frequencies. Therefore the cascading phenomena observed here
implies an altogether new behaviour on the gravitational side. Furthermore, as
described in~\cite{C-3,C-2,VanRaamsdonk:2008fp}, the full metric of the
corresponding spacetime can be obtained and the implications of this cascading
behaviour analyzed through suitable geometric quantities. We defer such tasks
to a forthcoming work. We close noting that this possible cascade behaviour,
first raised in~\cite{VanRaamsdonk:2008fp} and demonstrated here (and presented
in e.g.,~\cite{pirsa}) has now been observed in the Poincar\'{e} patch
case~\cite{chesler}.

Of course, another field theory `prediction' is that large AdS black holes in five and higher
dimensions will also exhibit turbulence but this chaotic behaviour will give
rise to a `standard' cascade to shorter wavelengths in these cases. From a
gravitational perspective, this apparently generic behaviour for `hot' horizons
in AdS is completely unexpected and calls for a better understanding within
gravity itself.

A step towards understanding the distinction between gravity in four and higher
dimensions can be obtained by recalling that enstrophy conservation is what
drives the hydrodynamics in 2+1 dimensions to exhibit the inverse cascade. One
can then exploit the duality to translate the enstrophy into geometrical
variables and to understand the implications of its conservation on the
gravitational side. That is, the conservation of enstrophy in the fluid
description implies a quasi-conserved quantity exists in the bulk gravity
theory. Further, as we have seen here, the system displays a rich dynamical
vortex configuration that merge towards a long lived state described by
relatively few, long wavelength vortices. Isolated vortices in conformal fluids
in 2+1 dimensions have been studied in~\cite{Evslin:2010ij} and their
properties identified. Again using the fluid/gravity correspondence, one could
produce a gravitational description of these quasi-stationary vortices. That
the understanding of these or other geometrical quantities might help shed new
light in turbulence phenomena is definitively an exciting prospect (see
e.g.,~\cite{Eling:2010vr}).

Certainly, the full gravitational description will naturally incorporate
dissipative contributions in the fluid flows. While we argued and
quantitatively demonstrated that this dissipation will not modify the essential
features of the turbulent behaviour at sufficiently high temperatures. In
Newtonian hydrodynamics, the onset of turbulence is discussed in terms of the
Reynolds number $Re$. The typical benchmark for the onset of turbulent flows is
that $Re$ have value of a few thousand. In relativistic hydrodynamics, the
latter may be estimated as \cite{foux}
\begin{equation}
Re \sim \frac{TL}{\eta/s} = 4\pi\,TL\,,
\label{reynolds}
\end{equation}
where $L$ is a characteristic length scale in the flow. In the last expression
above, we have substituted the celebrated holographic value $\eta/s=1/4\pi$
\cite{early0,early1,chris} --- we would have $\eta/s=\beta$ for the general
conformal fluid in \eqref{fluidexp2}. As this expression illustrates, we can
produce arbitrarily large values of $Re$ by increasing $T$ (while keeping the
system size fixed). Hence to observe turbulent behaviour, we are again
naturally pushed to the regime where the hydrodynamic gradient expansion works
well and our perfect fluid model becomes a good approximation. Of course, the
viscous effects will definitely modify the very long-time behaviour observed
here. For example, the conservation of the enstrophy is only true to first
order in the hydrodynamic gradient expansion.

Of course, another interesting extension of the present investigation would be
studying further the details of the turbulent cascade, in particular, the
Kolmogorov scaling exponents. Figure \ref{fig:torus_fourier2} illustrates some
preliminary results, which suggest that the `Newtonian' kinetic energy will
scale with the expected exponent of $-5/3$. However, various caveats must be
noted. First, a clean easy-to-distinguish Kolmogorov-type reasoning applies
to a situation where the turbulent flow is driven by an external force at some
high frequency (in the present case of 2+1 dimensions) and viscous dissipation also
damps the energy flow on much longer time scales. This scenario must be contrasted with the
freely-decaying turbulence (i.e., without any driving force) which
figure \ref{fig:torus_fourier2} describes. In freely-decaying turbulence determining
the intertial range is more delicate but in such range (which shrinks as time
proceeds) the $-5/3$ slope can be distinguished. Additionally, Kolmogorov's scaling
arguments are made in a Newtonian context and can at best be regarded as
approximate for the relativistic fluids studied here. Fortunately, the
appropriate exact scaling relations applicable for relativistic hydrodynamic
turbulence have been derived in \cite{foux}. The present simulations provide a
framework for further studying these relativistic relations. More generally,
however, the most exciting possibility would again be if a holographic
perspective could provide new insights into the issues surrounding these
turbulent cascades.

%
%
\noindent{\bf{\em Acknowledgments:}} It is a pleasure to thank P. Chesler, R.
Emparan, M. Kruczenski and T. Wiseman for interesting discussions.
This work was supported by NSERC through Discovery Grants and CIFAR (to LL and
RCM) and by CONICET, FONCYT and SeCyT-Univ. Nacional de Cordoba Grants (to FC
and OR). F.C. thanks Perimeter Institute's Visiting Graduate Fellows program
for hosting his stay at the Perimeter Institute, where parts of this work were
completed. Research at Perimeter Institute is supported through Industry Canada
and by the Province of Ontario through the Ministry of Research \& Innovation.
Computations were performed at SciNet.

\renewcommand{\theequation}{A-\arabic{equation}}
\setcounter{equation}{0}
\section*{Appendix A.}\label{appendixA}
For comparison purposes, we also consider scenarios related to the dual case of
an $AdS_4$ black brane solution
where the non-radial directions are compactified on a torus of size $D$.
In Fefferman-Graham coordinates, labeled by $(t,x,y)$,
the asymptotic metric is simply the flat one and the equations of motion can be straightforwardly
implemented.

\subsection{Stationary solutions and initial data}
Equilibrium configurations are given simply by constant flows at a given
(arbitrary) temperature $T$. For simplicity we consider a torus with domain
$[0,D]^2$. We then restrict to initial configurations with a flow along $x$,
i.e., $u^a =\delta_x^a u_0$, and introduce generic perturbations $\delta u^a$
to this flow. For concreteness, we describe here two particular cases where the
initial three-velocity is given by: case (A) with a perturbation of compact
support,
\begin{eqnarray}
u^a =
 \delta_x^a \left(u_0+ \delta u \sin(2 \pi y/3) \,y^2 \, (y-L)^2\right)
\end{eqnarray}
for $0\le y\le L$ and $u^a =\delta_x^a u_0$ otherwise; and case (B)
\begin{eqnarray}
u^a = \delta_x^a \left(u_0+ {\delta u} \sin(16 \pi y/D)\right)    \, .
\end{eqnarray}
We note however the qualitative features observed in all cases considered remain the same.

\subsection{Turbulence, cascading behaviour and temperature dependence}
We performed a series of numerical experiments adopting $u_0 = 0,\ 0.1$ and
$0.5$  and $\delta u = 0.01,\,\cdots,\, 0.05$ (i.e., perturbations from $2\%$
to $50\%$). and considered temperatures $T\in[1,10^3]$ with $L=10$ and typical
grid sizes of $[401,401]$ (though consistency of behaviour was checked with
grids $1.5$ and $2$ times better resolved).

\begin{figure}
\begin{center}
\epsfig{file=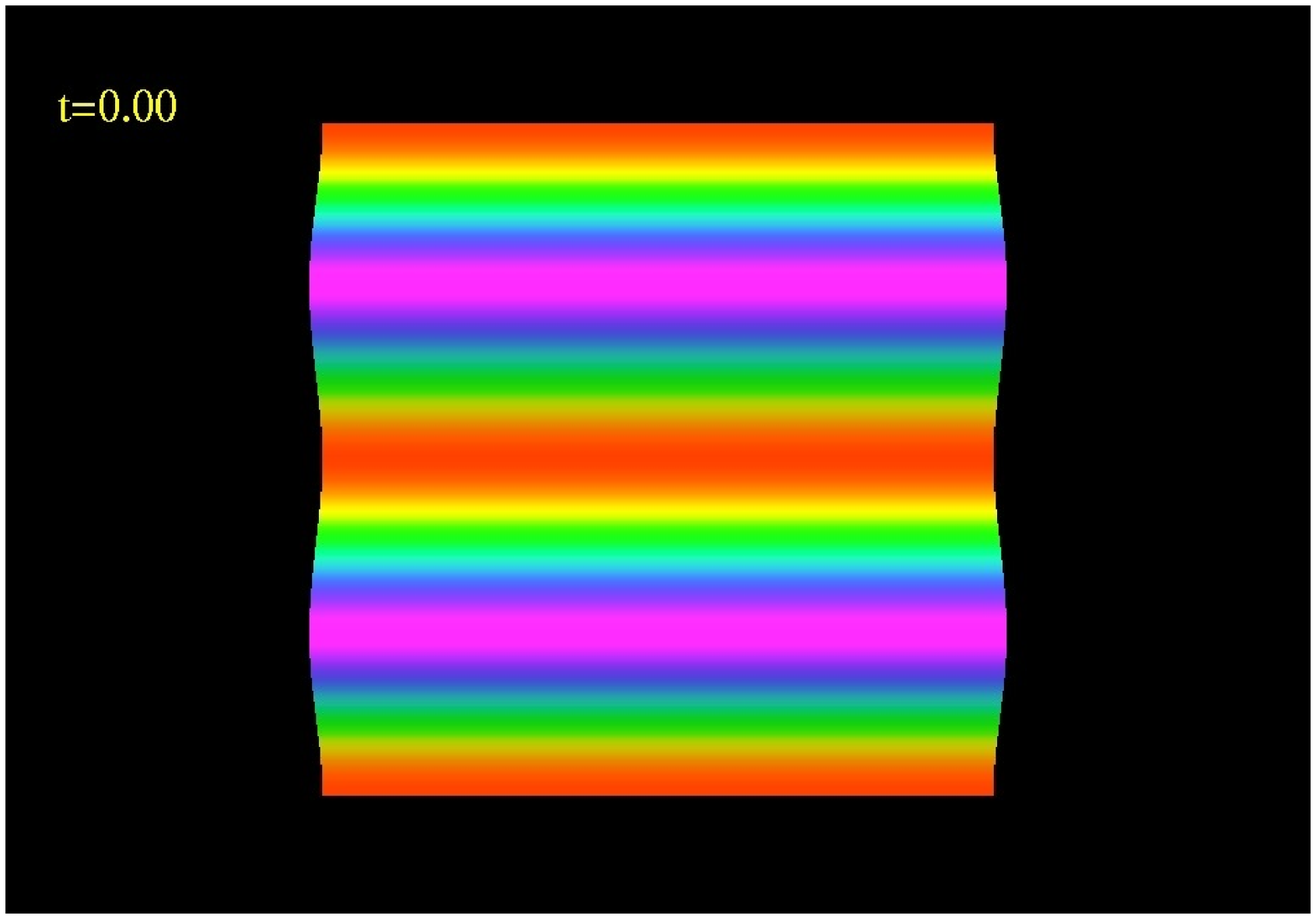,height=2.4cm}
\epsfig{file=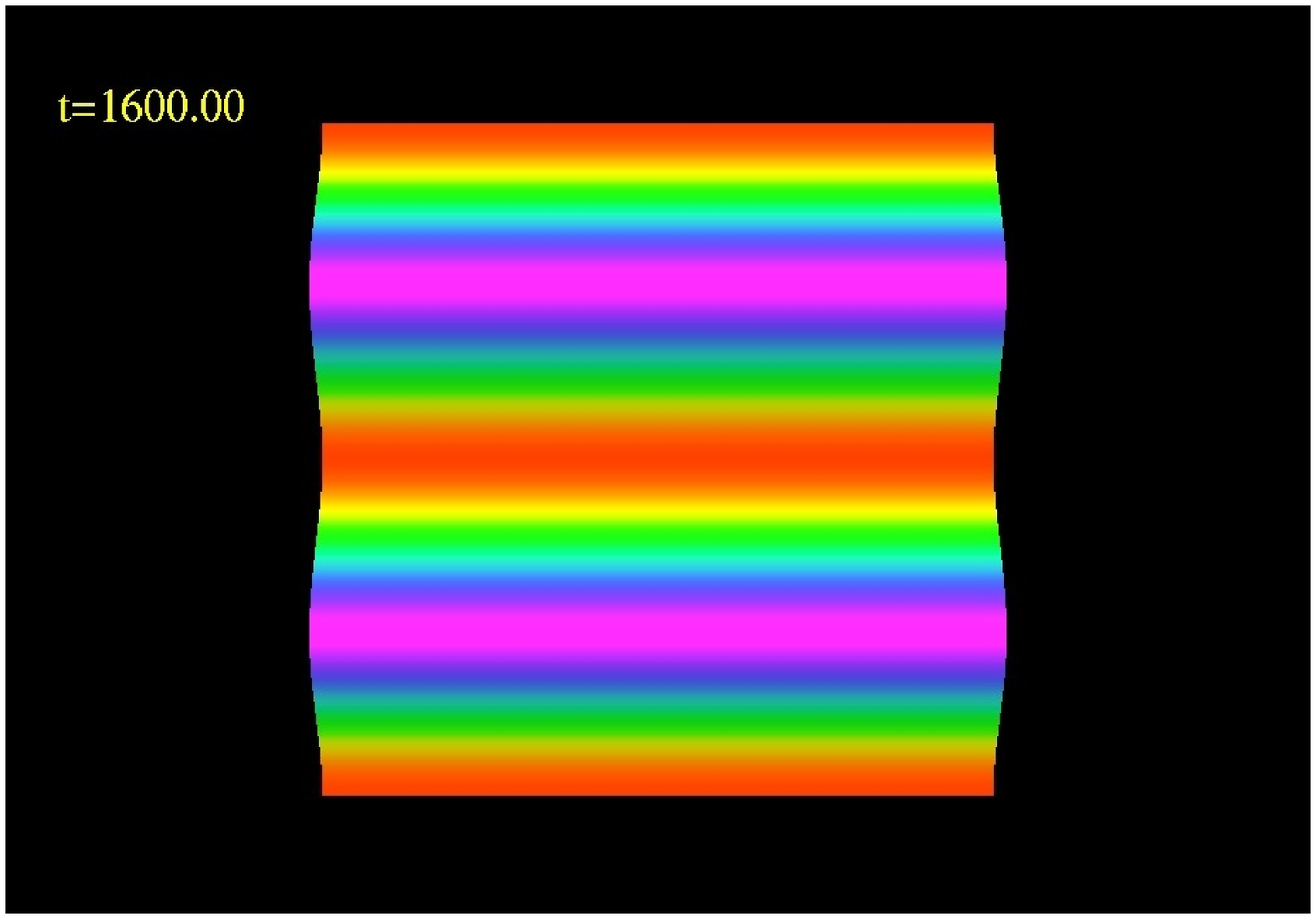,height=2.4cm} \\
\epsfig{file=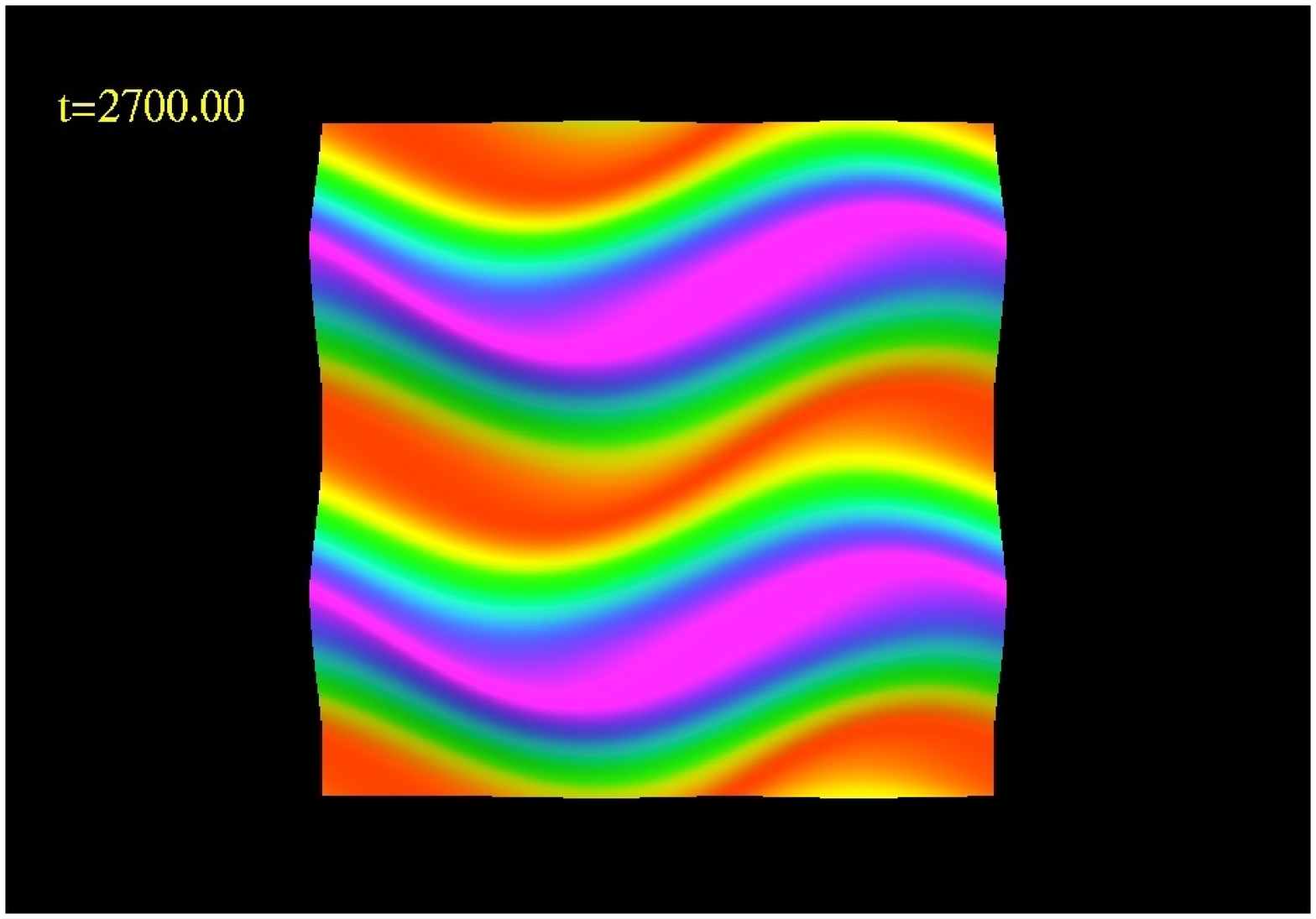,height=2.4cm}
\epsfig{file=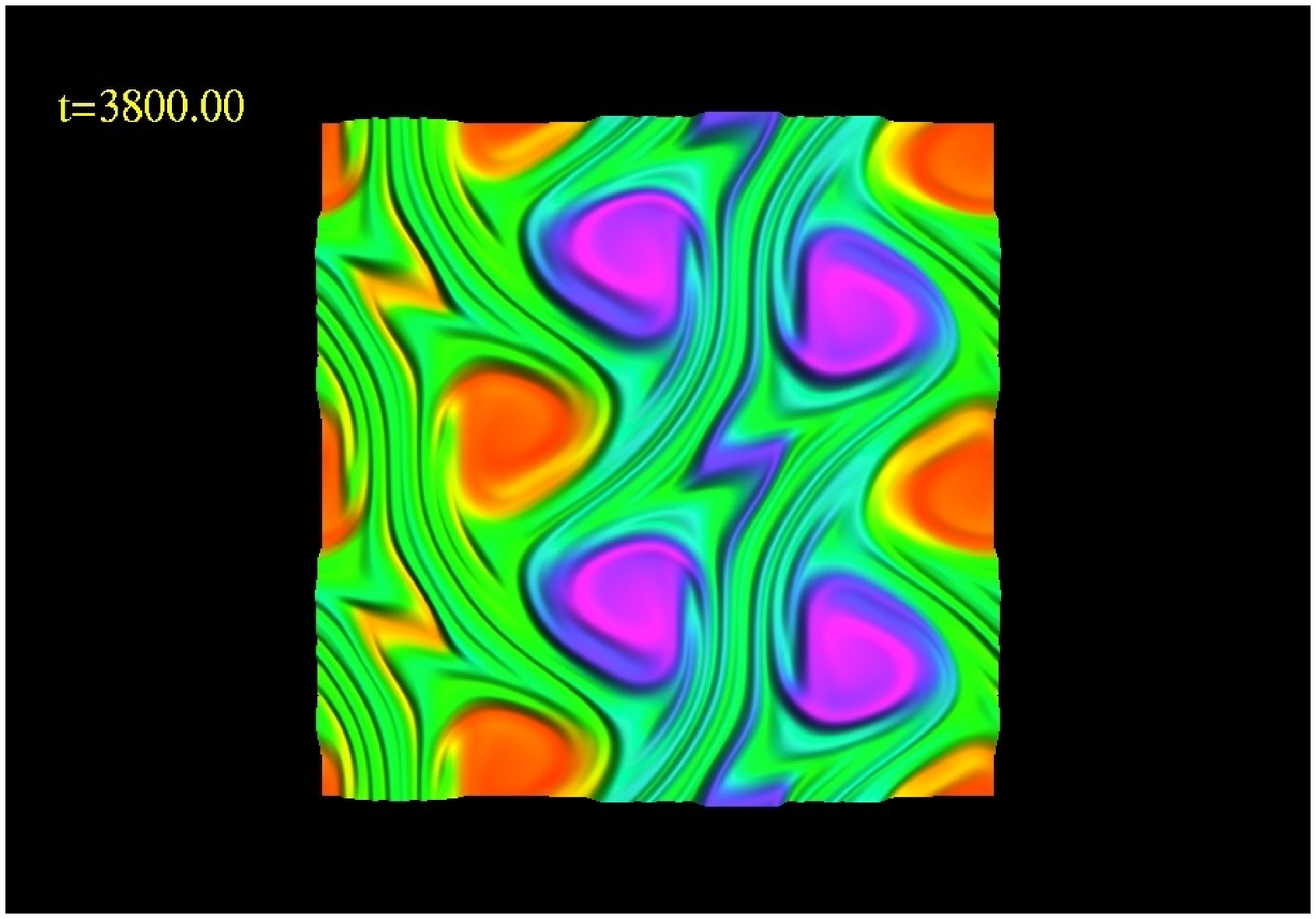,height=2.4cm} \\
\epsfig{file=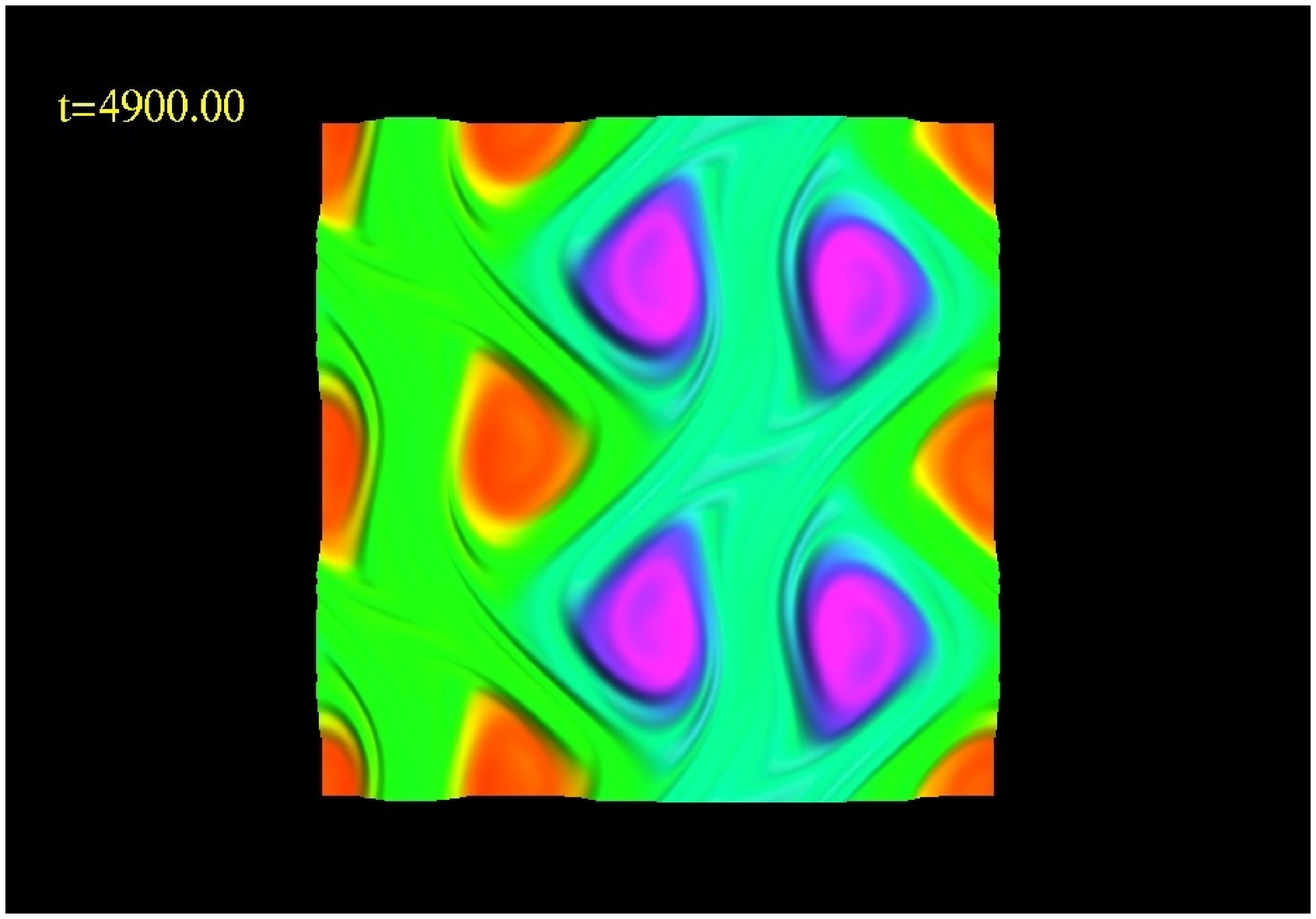,height=2.4cm}
\epsfig{file=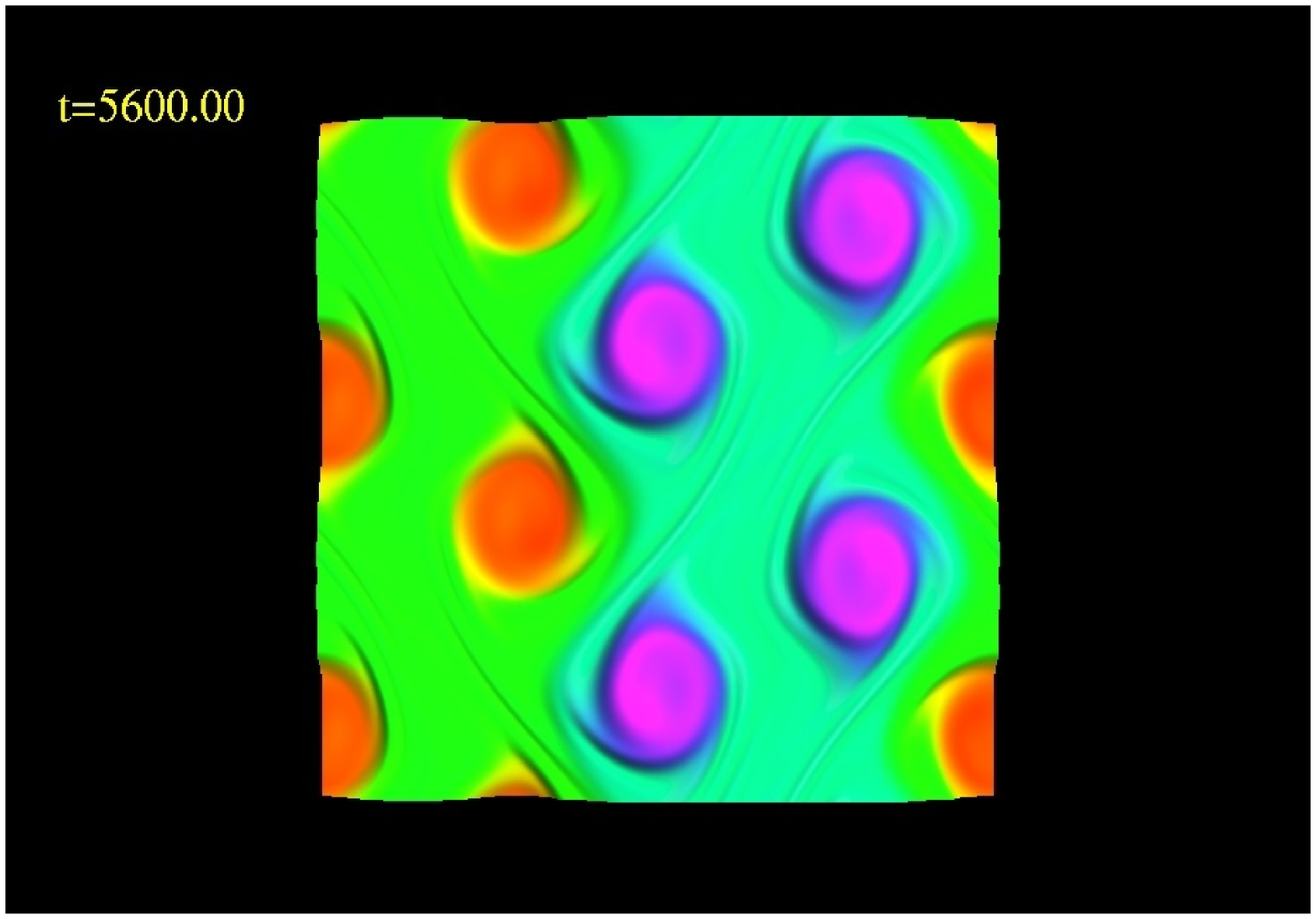,height=2.4cm}
\caption{Representative snapshots of the vorticity. As time progresses
the initial configuration is strongly disturbed by the formation of vortices.
As the dynamics continues, larger vortices are formed.} \label{fig:torus_vort}
\end{center}
\end{figure}

As in the case of the sphere, the dynamics display a turbulent behaviour
leading to the development of large vortices. As an illustrative example,
Figure~\ref{fig:torus_vort} displays the vorticity of the system at different
times for case (A) with $u_0 = 0.1$, $\delta u = 0.01$ and $T = 1$. Early on,
the behaviour is seemingly stationary but as turbulence develops the initial
symmetry is completely broken and vortices arise which grow as they merger
leading to a configuration described by long wavelengths.

\begin{figure}
\begin{center}
\epsfig{file=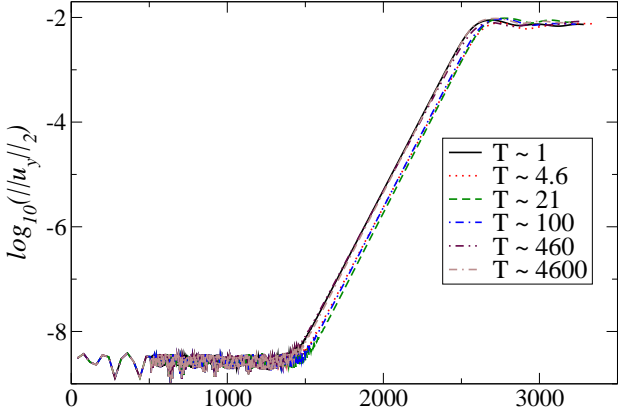,height=9.4cm,angle=-90}
\caption{Logarithm of the $L_2$ norm of $u^y$ for different temperatures.
Essentially the behaviour observed is unchanged with temperature.} \label{fig:torus_temp}
\end{center}
\end{figure}

As we have done in the $S^2$ case we also study the system's behaviour upon
variation of temperature to ensure terms neglected in our study do not
significantly affect the dynamics obtained. As in the previous case, a useful
proxy to monitor the evolution is the ($L_2$) norm of $u^y$ which, in the
absence of turbulent dynamics, would remain zero. Figure \ref{fig:torus_temp}
illustrates that the observed behaviour is essentially unchanged with
temperature.

Last, we also considered varying the size of the torus by increasing $D$ by
factors of $1.5$ and $2$ in case (A) and observed the same cascading behaviour.

\begin{figure}
\begin{center}
\epsfig{file=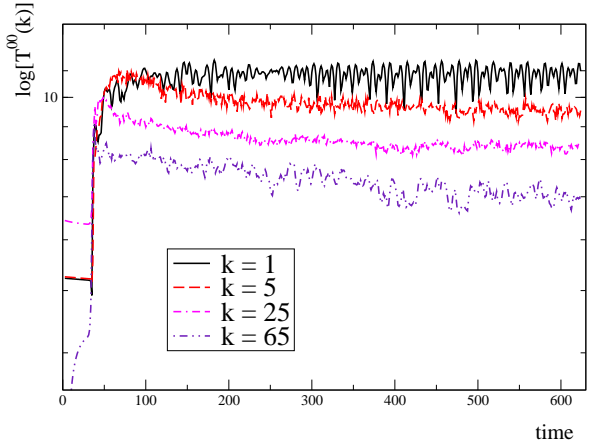,height=9.cm,angle=-90}
\epsfig{file=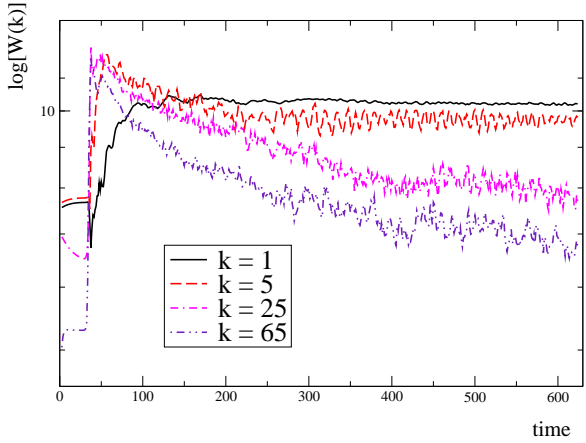,height=9.cm,angle=-90}
\caption{Fourier modes for energy density (top panel) and vorticity density (bottom panel). As
time progresses, a clear cascade to lower wave numbers is obtained.} \label{fig:torus_fourier}
\end{center}
\end{figure}

To study how the energy cascades as time progresses, we compute the Fourier
transform of energy density $T^{00}$ and the vorticity density $W^0$. Since our
computational grid is discretized by points $\{x_i^1,x_j^2\}$ (with
$i,j=1,\cdots,N$), we denote $\vec{n}=\{i,j\}$, and write
$T^{00}(\vec{x})=T^{00}(\vec{n})$ and the vorticity
$W^0(\vec{x})=W^0(\vec{n})$. The Fourier transform of these quantities is given
by
\begin{eqnarray}
\hat T^{00}(\vec{K}) &\,=\,& \sum_{\vec{n}} e^{-2\pi i \vec{K}.\vec{\frac{n}{N}}} T^{00}(\vec{n}) \,, \\
\hat W^0(\vec{K}) &\,=\,& \sum_{\vec{n}} e^{-2\pi i \vec{K}.\vec{\frac{n}{N}}} W^0(\vec{n}) \,.
\end{eqnarray}
Here $\vec{K}=\{K_1,K_2\}$ and $K_1,K_2 \in [0,N-1]$. Clearly, $\hat
T^{00}(\vec{K})$ and $\hat W^0(\vec{K})$ are also represented on a $N\times N$
grid. It is convenient to re-express the transformed quantities as functions of
$K=|\vec{K}|$. From the functions $\hat T^{00}(\vec{K})$ and $\hat
W^0(\vec{K})$, we define $T^{00}({K})$ and $W^0({K})$ as follows
\begin{eqnarray}
T^{00}({K}) &\,=\,& \sum_{i} |T^{00}(\vec{K}_i)|\,, \\
W^0({K}) &\,=\,& \sum_{i} |W^0(\vec{K}_i)|\,,
\end{eqnarray}
where the sums run over $\vec{K}_i$'s satisfying $K \leq |\vec{K_i}| < K+1$.
With this approach an effective wave-number grid of length $\sqrt{2}N$ is
obtained. Figure~\ref{fig:torus_fourier} illustrates a clear cascade to lower
wave-numbers as time proceeds. In the case of $T^{00}({K})$ the dominant mode
is given by $K=1$ while for the vorticity both $K=1$ and $K=2$ are of similar
magnitude. As in the $S^2$ case, this is expected as the vorticity field is
obtained by taking a single derivative which increases the mode content by one.

\begin{figure}
\begin{center}
\epsfig{file=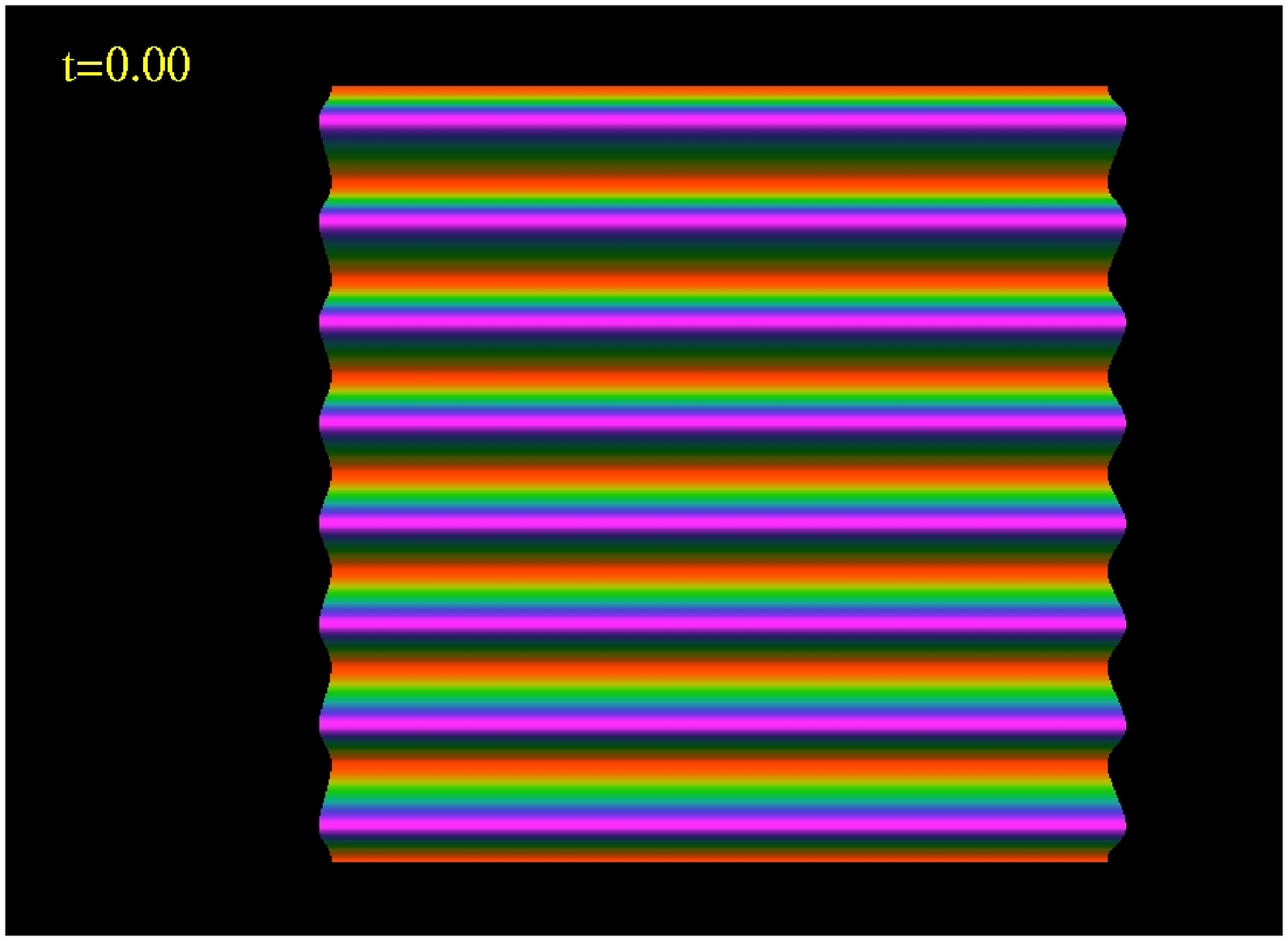,height=2.4cm}
\epsfig{file=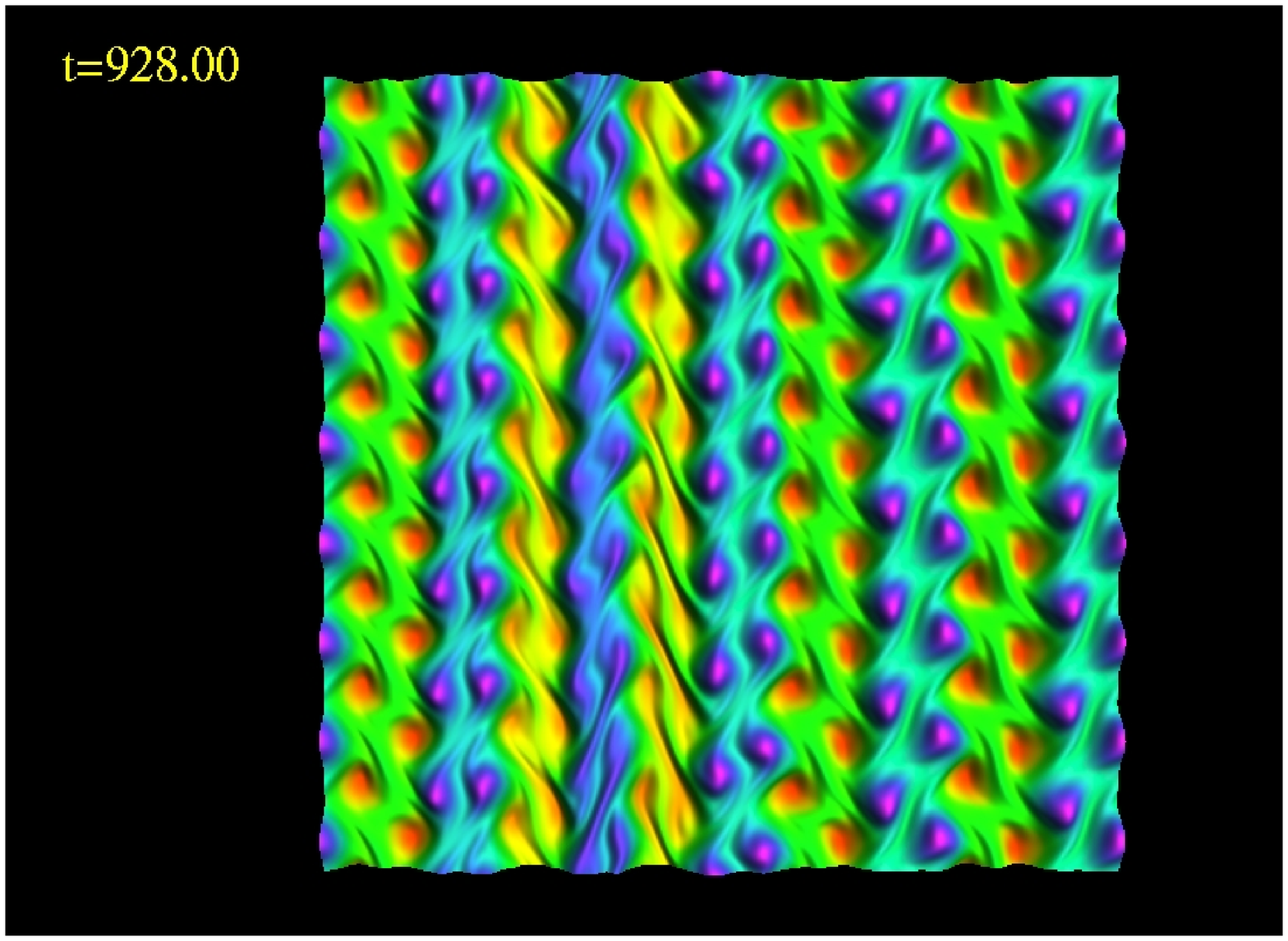,height=2.4cm} \\
\epsfig{file=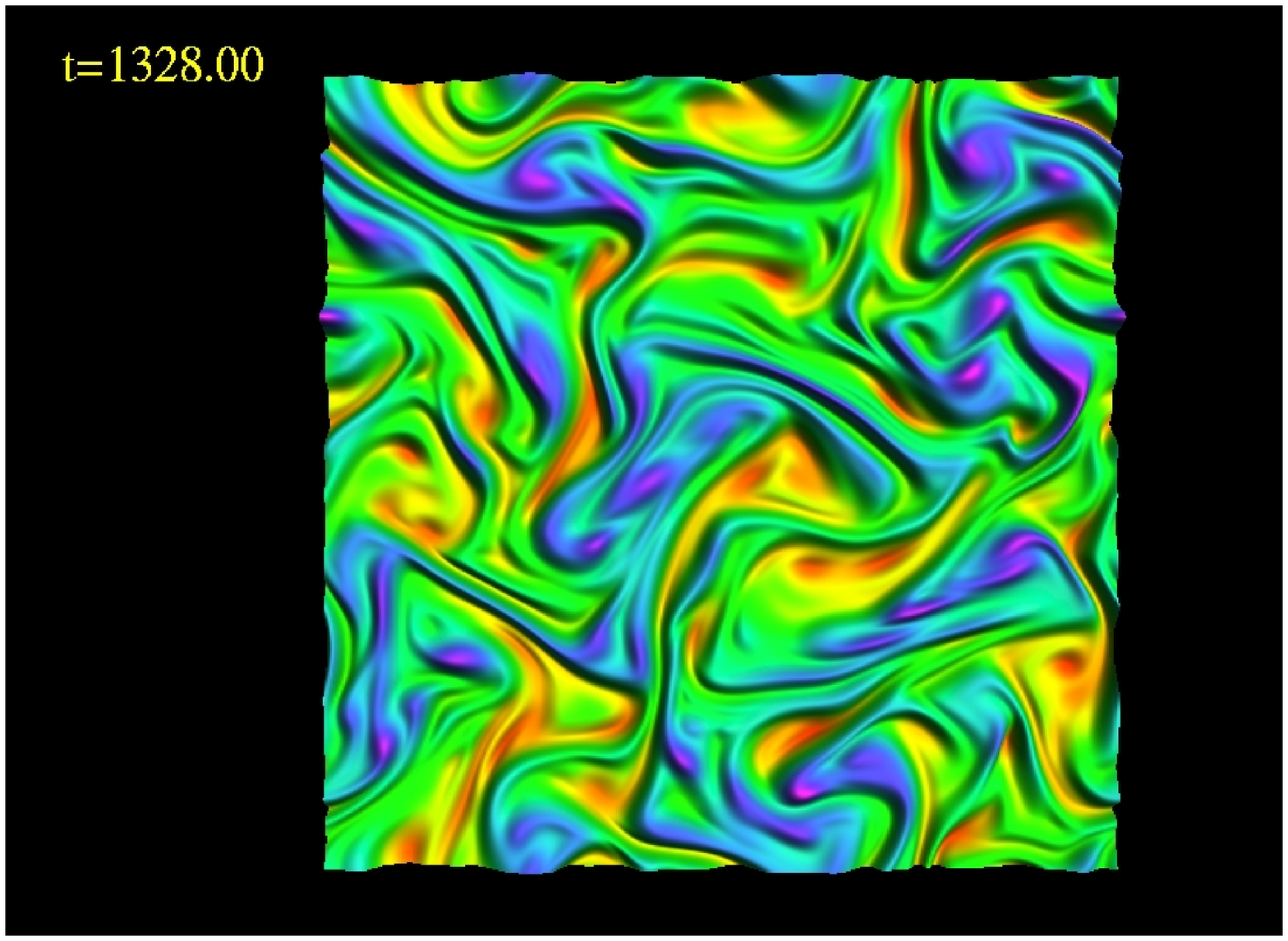,height=2.4cm}
\epsfig{file=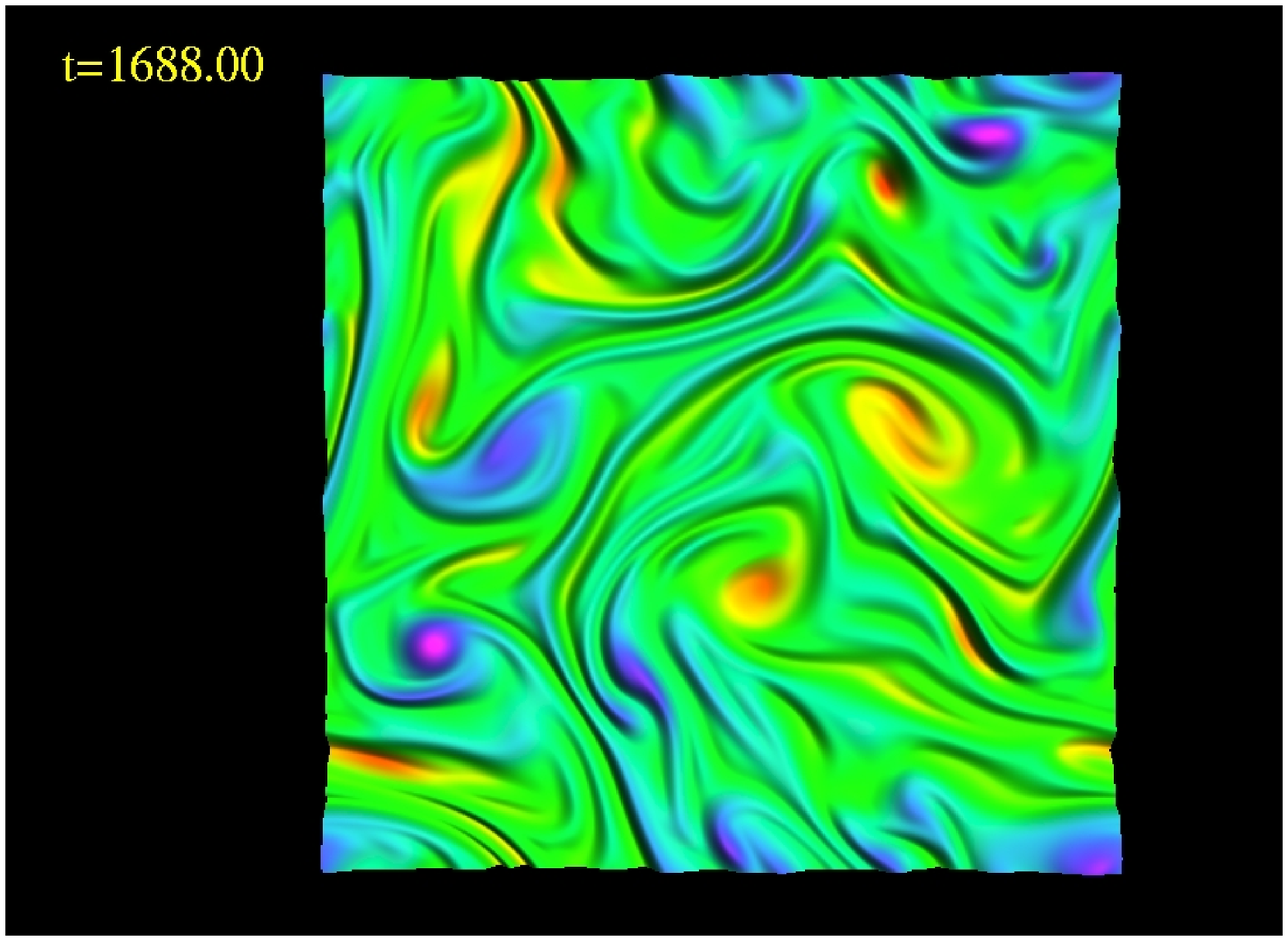,height=2.4cm} \\
\epsfig{file=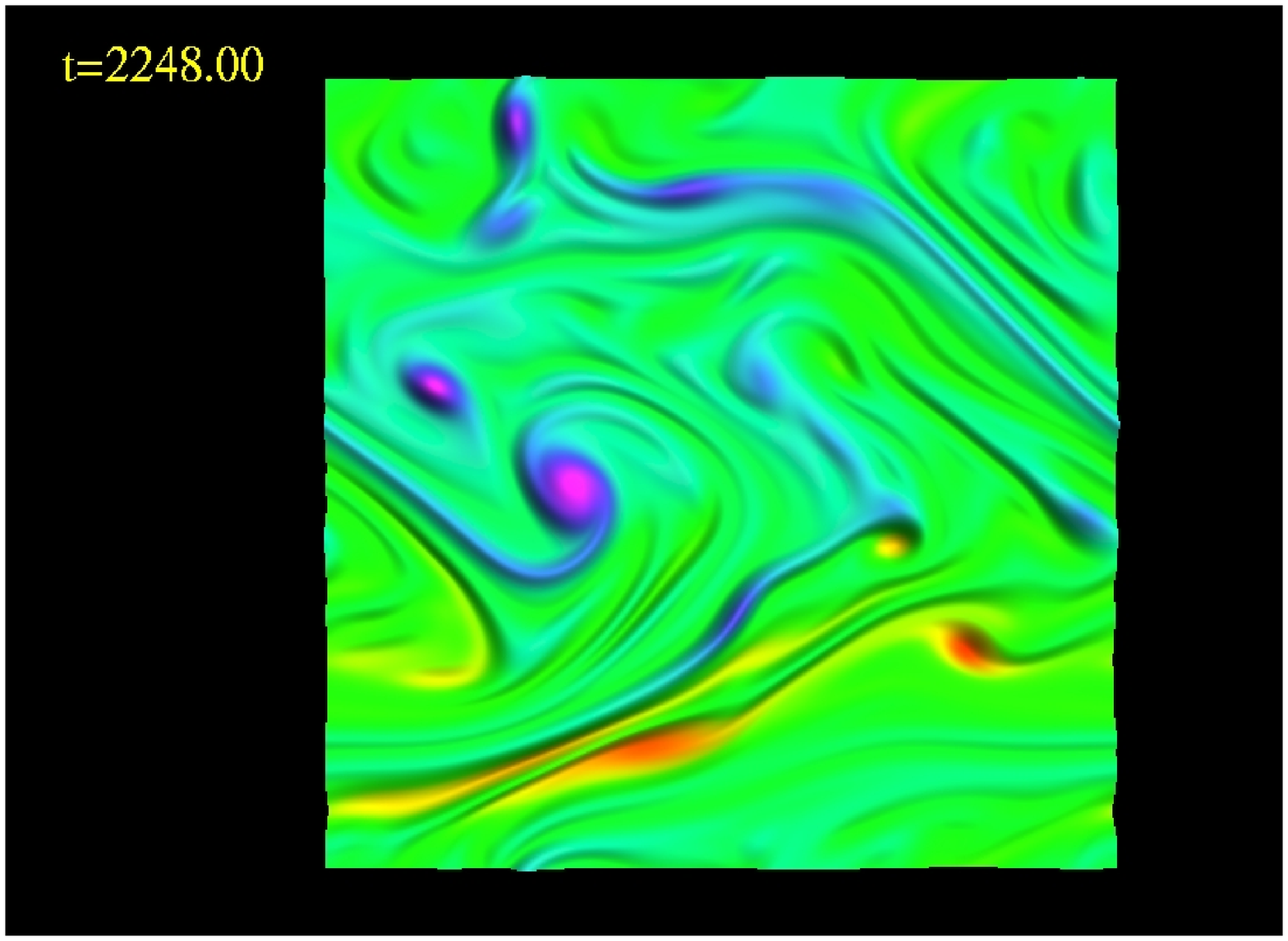,height=2.4cm}
\epsfig{file=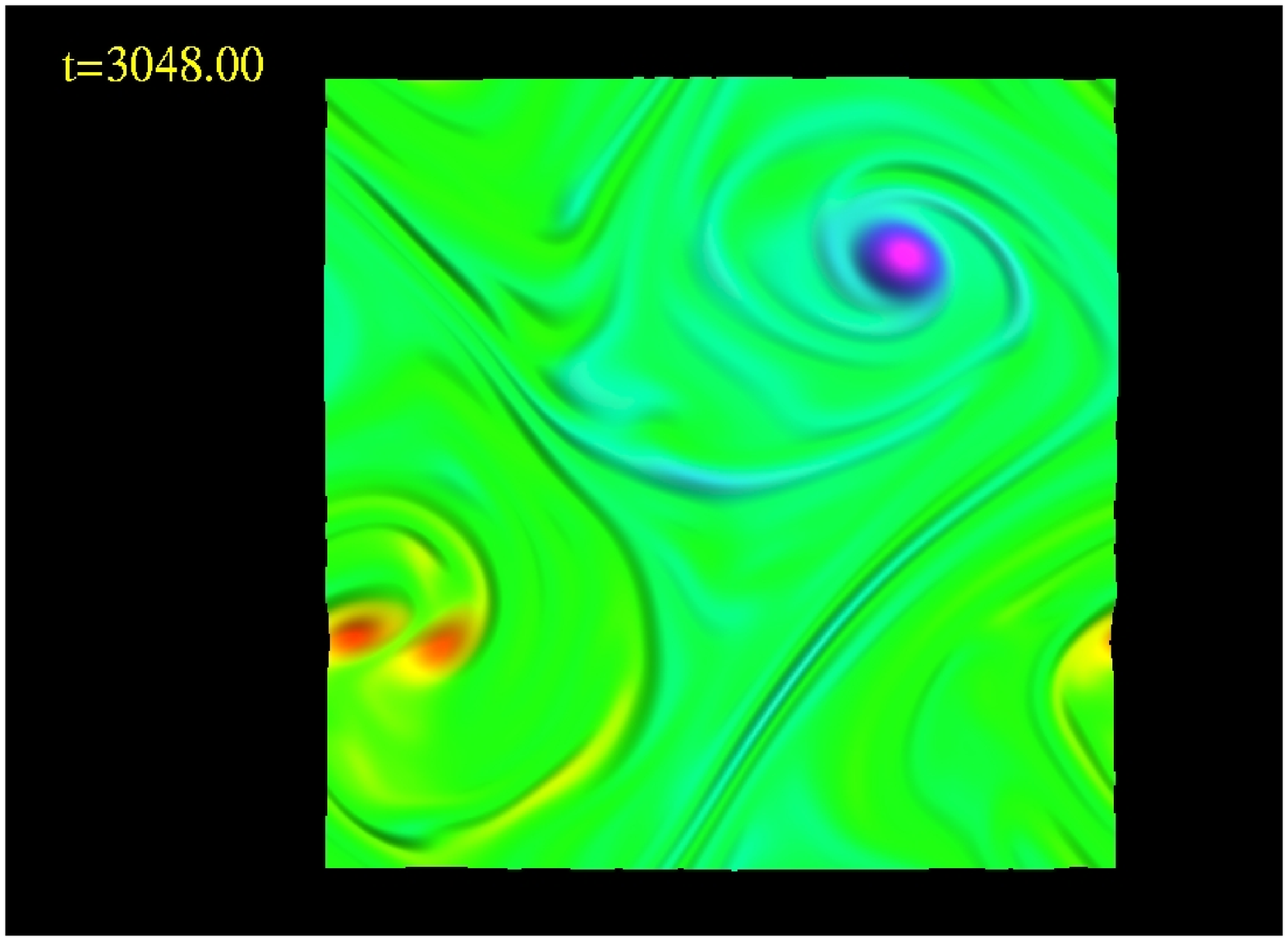,height=2.4cm}
\caption{Representative snapshots of the vorticity. Again, as time progresses
a turbulent behaviour is clearly displayed and an apparent cascade to lower wavelengths.} \label{fig:torus_vort2}
\end{center}
\end{figure}

\begin{figure}
\begin{center}
\epsfig{file=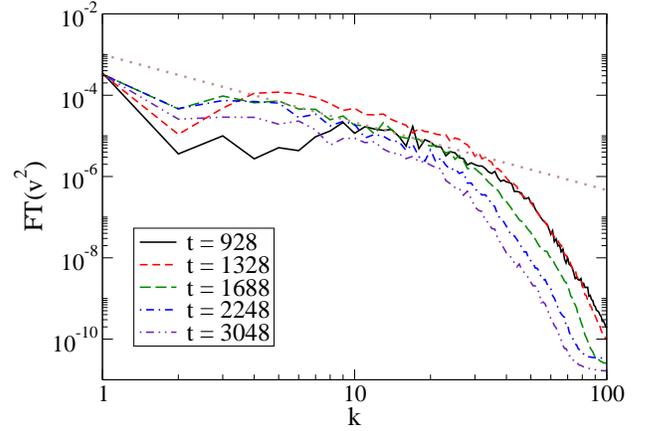,height=9.cm,angle=-90}
\caption{Fourier modes for the kinetic energy density for representative times.
As time progresses a cascading behaviour towards lower frequencies is manifested and
in an intermediate `inertial regime' with $k\in(8,400)$, the scaling is quite consistent
with a $-5/3$ slope (indicated by a dotted line in the figure).} \label{fig:torus_fourier2}
\end{center}
\end{figure}

It is interesting to monitor whether the observed behaviour is consistent with
the `standard' Kolmogorov expectation. To illustrate this, we compute the
Fourier transform of the `Newtonian' kinetic energy per unit mass $E=1/2 v^2$.
To make sense of this limit within our relativistic description, we chose $u_0
= 0$ and $\delta u = 0.03$, whose solution is such that $|v|$ is bounded by
$\simeq 0.04$ for all times, thus the fluid's motion stays far from
relativistic speeds. Figure \ref{fig:torus_vort2} shows the vorticity behaviour
for this case and \ref{fig:torus_fourier2} illustrates the Fourier
transformation of this kinetic energy for representative times. As time
proceeds the energy in higher frequency modes diminishes while the opposite
behaviour is observed for the lower ones. Additionally, at intermidiate
frequencies, the energy exhibits a behaviour with frequency consistent with a
slope of $-5/3$, the expected Kolmogorov exponent.

%
%
\bibliographystyle{unsrt}

\begin{thebibliography}{10}

\bibitem{Maldacena:1997re}
Juan~Martin Maldacena.
\newblock {The Large N limit of superconformal field theories and
  supergravity}.
\newblock {\em Adv.Theor.Math.Phys.}, 2:231--252, 1998.

\bibitem{magoo}
Ofer Aharony, Steven~S. Gubser, Juan~Martin Maldacena, Hirosi Ooguri, and Yaron
  Oz.
\newblock {Large N field theories, string theory and gravity}.
\newblock {\em Phys.Rept.}, 323:183--386, 2000.

\bibitem{C-3}
Sayantani Bhattacharyya, Veronika~E Hubeny, Shiraz Minwalla, and Mukund
  Rangamani.
\newblock {Nonlinear Fluid Dynamics from Gravity}.
\newblock {\em JHEP}, 0802:045, 2008.

\bibitem{C-4}
Sayantani Bhattacharyya, R.~Loganayagam, Ipsita Mandal, Shiraz Minwalla, and
  Ankit Sharma.
\newblock {Conformal Nonlinear Fluid Dynamics from Gravity in Arbitrary
  Dimensions}.
\newblock Technical Report TIFR/TH/08-38, 2008.

\bibitem{VanRaamsdonk:2008fp}
Mark {Van Raamsdonk}.
\newblock {Black Hole Dynamics From Atmospheric Science}.
\newblock {\em JHEP}, 0805:106, 2008.

\bibitem{early0}
P.~Kovtun, D.T. Son, and A.O. Starinets.
\newblock {Viscosity in strongly interacting quantum field theories from black
  hole physics}.
\newblock {\em Phys.Rev.Lett.}, 94:111601, 2005.

\bibitem{early1}
G.~Policastro, D.T. Son, and A.O. Starinets.
\newblock {The Shear viscosity of strongly coupled N=4 supersymmetric
  Yang-Mills plasma}.
\newblock {\em Phys.Rev.Lett.}, 87:081601, 2001.

\bibitem{Oz:2010wz}
Yaron Oz and Michael Rabinovich.
\newblock {The Penrose Inequality and the Fluid/Gravity Correspondence}.
\newblock {\em JHEP}, 1102:070, 2011.

\bibitem{Chesler:2010bi}
Paul~M. Chesler and Laurence~G. Yaffe.
\newblock {Holography and colliding gravitational shock waves in asymptotically
  AdS5 spacetime}.
\newblock {\em Phys.Rev.Lett.}, 106:021601, 2011.

\bibitem{Murata:2010dx}
Keiju Murata, Shunichiro Kinoshita, and Norihiro Tanahashi.
\newblock {Non-equilibrium Condensation Process in a Holographic
  Superconductor}.
\newblock {\em JHEP}, 1007:050, 2010.

\bibitem{CaronHuot:2011dr}
Simon Caron-Huot, Paul~M. Chesler, and Derek Teaney.
\newblock {Fluctuation, dissipation, and thermalization in non-equilibrium AdS5
  black hole geometries}.
\newblock {\em Phys.Rev.}, D84:026012, 2011.

\bibitem{Garfinkle:2011tc}
David Garfinkle, Leopoldo~A. Pando~Zayas, and Dori Reichmann.
\newblock {On Field Theory Thermalization from Gravitational Collapse}.
\newblock {\em JHEP}, 1202:119, 2012.

\bibitem{Bantilan:2012vu}
Hans Bantilan, Frans Pretorius, and Steven~S. Gubser.
\newblock {Simulation of Asymptotically AdS5 Spacetimes with a Generalized
  Harmonic Evolution Scheme}.
\newblock {\em Phys.Rev.}, D85:084038, 2012.

\bibitem{Buchel:2012gw}
Alex Buchel, Luis Lehner, and Robert~C. Myers.
\newblock {Thermal quenches in N=2* plasmas}.
\newblock 2012.

\bibitem{Chernicoff:2012gu}
Mariano Chernicoff, Daniel Fernandez, David Mateos, and Diego Trancanelli.
\newblock {Jet quenching in a strongly coupled anisotropic plasma}.
\newblock 2012.

\bibitem{2012PhRvE..85c6315C}
C.-k. {Chan}, D.~{Mitra}, and A.~{Brandenburg}.
\newblock {Dynamics of saturated energy condensation in two-dimensional
  turbulence}.
\newblock {\em \pre}, 85(3):036315, March 2012.

\bibitem{Bizon:2011gg}
Piotr Bizon and Andrzej Rostworowski.
\newblock {On weakly turbulent instability of anti-de Sitter space}.
\newblock {\em Phys.Rev.Lett.}, 107:031102, 2011.

\bibitem{movies}
Federico Carrasco, Luis Lehner, Robert Myers, Oscar Reula, and Ajay Singh.
\newblock {Conformal Fluids and Turbulent Behaviour in 2+1 Dimensions}.
\newblock Technical Report
  http://spaces.perimeterinstitute.ca/2d-turbulence/content/2d-turbulence,
  2012.

\bibitem{torrid}
P.A. Davidson.
\newblock {\em Turbulence: An Introduction for Scientists and Engineers}.
\newblock Oxford University Press, USA, 2004.

\bibitem{Sulem}
{Rose, H.A.} and {Sulem, P.L.}
\newblock {Fully developed turbulence and statistical mechanics}.
\newblock {\em J. Phys. France}, 39(5):441--484, 1978.

\bibitem{easy}
Krzysztof Gawedzki.
\newblock {Easy turbulence}.
\newblock Technical Report IHES/P/99/56, August 1999.

\bibitem{Kraichnan}
R.H. Kraichnan.
\newblock {Inertial ranges in two-dimensional turbulence}.
\newblock {\em Phys. Fluids}, 10:1417--1423, 1967.

\bibitem{Batchelor}
G.K. Batchelor.
\newblock {High Speed Computing in Fluid Dynamics}.
\newblock {\em Phys. Fluids}, 12:II--233, 1969.

\bibitem{Mcwilliams}
James~C. Mcwilliams.
\newblock {The emergence of isolated coherent vortices in turbulent flow}.
\newblock {\em Journal of Fluid Mechanics}, 146:21--43, 1984.

\bibitem{Borue}
Vadim Borue.
\newblock {Inverse energy cascade in stationary two-dimensional homogeneous
  turbulence}.
\newblock {\em Phys. Rev. Lett.}, 72:1475--1478, Mar 1994.

\bibitem{Benzi-2}
R~Benzi, G~Paladin, S~Patarnello, P~Santangelo, and A~Vulpiani.
\newblock {Intermittency and coherent structures in two-dimensional
  turbulence}.
\newblock {\em Journal of Physics A: Mathematical and General}, 19(18):3771,
  1986.

\bibitem{Muller:1967zza}
Ingo Muller.
\newblock {Zum Paradoxon der Warmeleitungstheorie}.
\newblock {\em Z.Phys.}, 198:329--344, 1967.

\bibitem{Israel:1979wp}
W.~Israel and J.M. Stewart.
\newblock {Transient relativistic thermodynamics and kinetic theory}.
\newblock {\em Annals Phys.}, 118:341--372, 1979.

\bibitem{Geroch1990}
Robert Geroch and Lee Lindblom.
\newblock {Dissipative relativistic fluid theories of divergence type}.
\newblock {\em Phys. Rev. D}, 41:1855--1861, Mar 1990.

\bibitem{Geroch1991}
Robert Geroch and Lee Lindblom.
\newblock {Causal theories of dissipative relativistic fluids}.
\newblock {\em Annals of Physics}, 207(2):394--416, 1991.

\bibitem{Buchel:2009tt}
Alex Buchel and Robert~C. Myers.
\newblock {Causality of Holographic Hydrodynamics}.
\newblock {\em JHEP}, 0908:016, 2009.

\bibitem{Baier:2007ix}
Rudolf Baier, Paul Romatschke, Dam~Thanh Son, Andrei~O. Starinets, and
  Mikhail~A. Stephanov.
\newblock {Relativistic viscous hydrodynamics, conformal invariance, and
  holography}.
\newblock {\em JHEP}, 0804:100, 2008.

\bibitem{paul}
Paul Romatschke.
\newblock {Relativistic Viscous Fluid Dynamics and Non-Equilibrium Entropy}.
\newblock {\em Class.Quant.Grav.}, 27:025006, 2010.

\bibitem{Kreiss}
B.~Gustafsson J.~Oliger {H. Kreiss}.
\newblock {\em {Time Dependent Problems And Difference Methods}}.
\newblock John Wiley and Sons, Inc., 1995.

\bibitem{Reula}
Gioel Calabrese, Luis Lehner, Oscar Reula, Olivier Sarbach, and Manuel Tiglio.
\newblock {Summation by parts and dissipation for domains with excised
  regions}.
\newblock Technical Report LSU-REL-080103, 2004.

\bibitem{C-2}
Sayantani Bhattacharyya, Subhaneil Lahiri, R.~Loganayagam, and Shiraz Minwalla.
\newblock {Large rotating AdS black holes from fluid mechanics}.
\newblock {\em JHEP}, 0809:054, 2008.

\bibitem{Gibbons200549}
G.W. Gibbons, H.~L{\"u}, Don~N. Page, and C.N. Pope.
\newblock The general kerr--de sitter metrics in all dimensions.
\newblock {\em Journal of Geometry and Physics}, 53(1):49--73, 2005.

\bibitem{Carter}
Brandon Carter.
\newblock {Covariant theory of conductivity in ideal fluid or solid media}.
\newblock {\em Lecture Notes in Mathematics}, 1385:1--64, 1989.

\bibitem{Leco}
Luis Lehner, Oscar Reula, and Manuel Tiglio.
\newblock {Multi-block simulations in general relativity: high order
  discretizations, numerical stability, and applications}.
\newblock {\em Class.Quant.Grav.}, 22:5283--5322, 2005.

\bibitem{parisireula}
Florencia Parisi and Oscar Reula.
\newblock in preparation.
\newblock 2012.

\bibitem{Carpenter1999}
Mark~H. Carpenter, Jan Nordstr{\"o}m, and David Gottlieb.
\newblock {A Stable and Conservative Interface Treatment of Arbitrary Spatial
  Accuracy}.
\newblock {\em Journal of Computational Physics}, 148(2):341--365, 1999.

\bibitem{Carpenter2001}
Jan Nordstr{\"o}m and Mark~H. Carpenter.
\newblock {High-Order Finite Difference Methods, Multidimensional Linear
  Problems, and Curvilinear Coordinates}.
\newblock {\em Journal of Computational Physics}, 173(1):149--174, 2001.

\bibitem{Winstanley:2001nx}
Elizabeth Winstanley.
\newblock {On classical superradiance in Kerr-Newman - anti-de Sitter black
  holes}.
\newblock {\em Phys.Rev.}, D64:104010, 2001.

\bibitem{Evslin:2012zn} 
  J.~Evslin,
\newblock {Hydrodynamic Vortices and their Gravity Duals}.
 \newblock {\em  Fortsch.\ Phys.\ } 60:1005, 2012.
  [arXiv:1201.6442 [hep-th]].

\bibitem{Dias:2011ss}
Oscar~J.C. Dias, Gary~T. Horowitz, and Jorge~E. Santos.
\newblock {Gravitational Turbulent Instability of Anti-de Sitter Space}.
\newblock {\em Class.Quant.Grav.}, 29:194002, 2012.

\bibitem{pirsa}
Federico Carrasco, Luis Lehner, Robert Myers, Oscar Reula, and Ajay Singh.
\newblock {Conformal Fluids and Turbulent Behaviour in 2+1 Dimensions}.
\newblock Technical Report http://pirsa.org/12060025/, 2012.

\bibitem{chesler}
Paul Chesler.
\newblock in progress.
\newblock 2012.

\bibitem{Evslin:2010ij}
Jarah Evslin and Chethan Krishnan.
\newblock {Vortices in (2+1)d Conformal Fluids}.
\newblock {\em JHEP}, 1010:028, 2010.

\bibitem{Eling:2010vr}
Christopher Eling, Itzhak Fouxon, and Yaron Oz.
\newblock {Gravity and a Geometrization of Turbulence: An Intriguing
  Correspondence}.
\newblock 2010.

\bibitem{foux}
Itzhak Fouxon and Yaron Oz.
\newblock {Exact Scaling Relations In Relativistic Hydrodynamic Turbulence}.
\newblock {\em Phys.Lett.}, B694:261--264, 2010.

\bibitem{chris}
Christopher~P. Herzog.
\newblock {The Hydrodynamics of M theory}.
\newblock {\em JHEP}, 0212:026, 2002.

\bibitem{Liu:2010jg}
Xiao Liu and Yaron Oz.
\newblock {Shocks and Universal Statistics in (1+1)-Dimensional Relativistic
  Turbulence}.
\newblock {\em JHEP}, 1103:006, 2011.

\bibitem{Bredberg:2011jq}
Irene Bredberg, Cynthia Keeler, Vyacheslav Lysov, and Andrew Strominger.
\newblock {From Navier-Stokes To Einstein}.
\newblock {\em JHEP}, 1207:146, 2012.

\end{thebibliography}

\end{document}